\documentclass[aps,twocolumn,prd,superscriptaddress,showpacs,preprintnumbers,
amsmath,amssymb,nofootinbib] {revtex4-1}

\usepackage{color}
\usepackage{graphicx}
\usepackage{amsmath, amssymb}
\usepackage{amsfonts}
\usepackage{dcolumn}
\usepackage{subfigure}
\usepackage{hyperref}

\newcommand{\Tr}{\ensuremath{\operatorname{Tr}}}

\def\eq#1{Eq.~(\ref{#1})}
\def\eqref#1{(\ref{#1})}
\def\fig#1{Fig.~\ref{#1}}
\def\tab#1{Tab.~\ref{#1}}

\def\sec#1{Sec.~\ref{#1}}
\newcommand{\Phibar}{\ensuremath{\bar{\Phi}}}

\def\dr{{D\!\llap{/}}\,}

\def\0#1#2{\frac{#1}{#2}}

\newcommand{\mrm}[1]{\mathrm{#1}}

\newcommand{\sx}{\sigma_{x}}
\newcommand{\sy}{\sigma_{y}}
\def\pd{\partial}

\graphicspath{{./}}

\begin{document}

\title{Thermodynamics of QCD at vanishing density}

\author{Tina Katharina Herbst}
\affiliation{Institut f\"ur Theoretische Physik,
Universit\"at Heidelberg, 
Philosophenweg 16, D-69120 Heidelberg, Germany}

\author{Mario Mitter}
\affiliation{Institut f\"ur Theoretische Physik, Universit\"at Heidelberg, 
Philosophenweg 16, D-69120 Heidelberg, Germany}
\affiliation{Institut f\"ur Theoretische Physik, Goethe-Universit\"at Frankfurt, 
Max-von-Laue-Stra{\ss}e 1 , D-60438 Frankfurt/Main, Germany}

\author{Jan M. Pawlowski}
\affiliation{Institut f\"ur Theoretische Physik, Universit\"at Heidelberg, 
Philosophenweg 16, D-69120 Heidelberg, Germany}
\affiliation{ExtreMe Matter Institute EMMI, GSI Helmholtzzentrum f\"ur 
Schwerionenforschung mbH, Planckstra{\ss}e 1, D-64291 Darmstadt, Germany}

\author{Bernd-Jochen Schaefer}
\affiliation{Institut f\"{u}r Physik, Karl-Franzens-Universit\"{a}t,
Universit\"atsplatz 5, A-8010 Graz, Austria}
\affiliation{Institut f\"{u}r Theoretische Physik, Justus-Liebig-Universit\"at,
Heinrich-Buff-Ring 16, D-35392 Giessen, Germany}

\author{Rainer Stiele}
\affiliation{Institut f\"ur Theoretische Physik, Universit\"at Heidelberg, 
Philosophenweg 16, D-69120 Heidelberg, Germany}
\affiliation{ExtreMe Matter Institute EMMI, GSI Helmholtzzentrum f\"ur 
Schwerionenforschung mbH, Planckstra{\ss}e 1, D-64291 Darmstadt, Germany}


\begin{abstract}
  \noindent We study the phase structure of QCD at finite temperature
  within a Polyakov-loop extended quark-meson model. Such a model
  describes the chiral as well as the confinement-deconfinement
  dynamics. In the present investigation, based on the approach and
  results put forward in
  \cite{Braun:2009gm,Herbst:2013ail,Haas:2013qwp, Mitter:2013fxa},
  both, matter as well as glue fluctuations are included. We present
  results for the order parameters as well as some thermodynamic
  observables and find very good agreement with recent results from
  lattice QCD.
\end{abstract}

\maketitle

\section{Introduction}\label{sec:intro}

The study of strongly interacting matter under extreme conditions is a
very active field of research. Experiments conducted at CERN, RHIC and the
future FAIR and NICA facilities aim at probing the phase structure of Quantum
Chromodynamics (QCD).

From the theoretical side, calculating the phase structure from first
principles is a hard task which requires the use of non-perturbative
methods. Over the recent years a lot of progress has been made in this
direction. In particular it has been shown that, apart from lattice
QCD, also continuum methods, such as the Functional Renormalisation
Group (FRG) \cite{Litim:1998nf, Berges:2000ew, Pawlowski:2005xe,
  Gies:2006wv, Schaefer:2006sr,Pawlowski:2010ht,Braun:2011pp,
  vonSmekal:2012vx} are well suited to study the QCD phase
diagram. This has been demonstrated in e.g.~\cite{Braun:2007bx,
  Braun:2009gm,Fischer:2012vc, Fister:2013bh, Fischer:2013eca,
  Haas:2013qwp} at vanishing as well as finite temperature and
chemical potential. Complementary to first-principles studies,
low-energy QCD has been studied successfully within effective
models. Especially the use of Polyakov-loop extended chiral models
makes it possible to study the interrelation of the chiral and
deconfinement phase transitions, e.g.~\cite{Meisinger:1995ih,
  Pisarski:2000eq, Fukushima:2003fw, Megias:2004hj, Ratti:2005jh,
  Mukherjee:2006hq, Roessner:2006xn, Sasaki:2006ww, Schaefer:2007pw,
  Fraga:2007un, Fukushima:2008wg, Sakai:2008py, Herbst:2010rf,
  Skokov:2010uh, Skokov:2010wb, Schaefer:2011ex, Braun:2011fw,
  Kamikado:2012bt, Braun:2012zq, Fukushima:2012qa, Mintz:2012mz,
  Herbst:2013ail, Haas:2013qwp, Stiele:2013pma, Strodthoff:2013cua}.
However, the confinement sector in these models is not fully
constrained, resulting in various parametrisations of the
corresponding order-parameter potential, the glue or Polyakov-loop
potential.  Furthermore, the important unquenching effects on the glue
potential are usually not included. Ideally, this potential is derived
from QCD directly, leaving no ambiguity. This has recently been
accomplished with the FRG for two-flavour QCD in the chiral limit
\cite{Braun:2009gm} and for $2+1$ flavours in \cite{Fischer:2013eca}
and puts us in the position to make use of these results to improve
the effective description of the gauge sector. In summary, these
effective models can be systematically improved towards full QCD,
using input from the lattice and other first-principles studies, see
e.g.~\cite{Braun:2009gm, Pawlowski:2010ht, Herbst:2013ail,
  Haas:2013qwp}. In \cite{Haas:2013qwp, Stiele:2013gra} this approach
has already been tested in a mean-field approximation.

In the present work we aim at quantitative results for the thermodynamics of
QCD. To achieve this goal, we combine the previous efforts of
\cite{Herbst:2013ail, Mitter:2013fxa} and include quantum and
thermal fluctuations with the FRG in an effective Polyakov--quark-meson (PQM)
model with $2+1$ flavours. Furthermore, we apply the augmentation of the gauge
sector by QCD results as in \cite{Haas:2013qwp}. In combination, this gives us a
good handle on the chiral and confinement-deconfinement transitions and
thermodynamics of QCD. 

This work is structured as follows: In Sec.~\ref{sec:PQM} we briefly review the
FRG approach to QCD and its connection to low-energy effective models. In 
particular, we discuss how to augment low-energy effective models with
first-principles results from QCD. In Sec.~\ref{sec:PQM2p1} we provide the
details of our truncation and present the resulting flow equation in
Sec.~\ref{subsec:PQMflow}.
Results for the order parameters and thermodynamic observables for $2+1$
flavours are presented in Sec.~\ref{subsec:crossover} and
Sec.~\ref{subsec:TD2+1}, respectively. Sec.~\ref{subsec:TD2} contains our
prediction for the thermodynamics in the two-flavour case.
Concluding remarks and a summary are presented in Sec.~\ref{sec:conclusion}. We
discuss the dependence of our results on various parameters in the appendix.

\section{Functional Renormalisation Group Approach to Low-Energy
QCD}\label{sec:PQM}

The mapping of QCD degrees of freedom to low-energy effective models is
discussed in depth in, e.g.~\cite{Braun:2009gm, Pawlowski:2010ht,
Herbst:2013ail, Haas:2013qwp}. Here, we only briefly recapitulate the main
points.
\begin{figure}
  \includegraphics[width=\columnwidth]{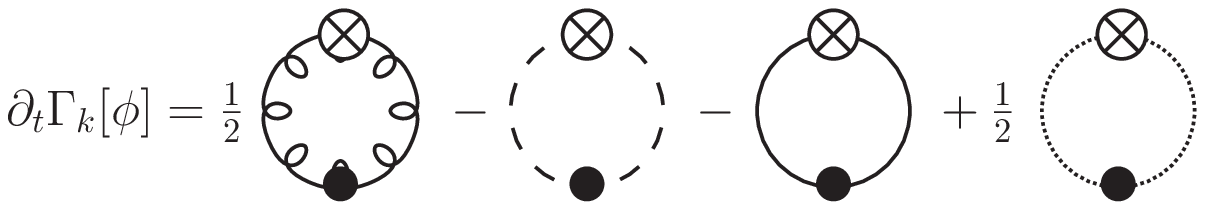}
  \caption{Partially hadronised version of the FRG flow for QCD. The
    loops denote the gluon, ghost, quark and meson contributions,
    respectively. The crosses mark the FRG regulator term.}
  \label{fig:QCDflow}
\end{figure}

\fig{fig:QCDflow} shows the pictorial representation of the FRG flow
of QCD, where the first two loops represent the gluon and ghost
contributions, respectively, whereas the third loop denotes the quark
degrees of freedom.  The fourth loop corresponds to mesonic degrees of
freedom which have been introduced via the dynamical hadronisation
technique \cite{Gies:2001nw, Gies:2002hq, Pawlowski:2005xe,
Floerchinger:2009uf}. 

It is well-established that the ghost-gluon sector decouples from the
matter dynamics below the chiral and deconfinement temperatures, see
e.g.~\cite{Fischer:2009ah}. In terms of the flow equation,
\fig{fig:QCDflow}, this means that in this regime we are only left
with the dynamical matter sector given by the last two loops,
explicitly
\begin{eqnarray}\label{eq:matterflow}
  \partial_t \Delta\Gamma_k[\bar A; \phi] & = & - \Tr \left(G_q[\bar A; \phi]
    \partial_t R_q\right) \nonumber\\
    & & +\ \frac{1}{2}\Tr \left(G_H[\bar A; \phi]\partial_t
  R_H\right)\,.
\end{eqnarray}
The full field content is collected in $\phi~=~(a,\ c,\ \bar c,\ q,\
\bar q,\ H)$\,. In the non-perturbative domain of QCD
the spectrum is gapped and only light constituent quarks ($q,\bar q$) and the
corresponding hadrons ($H$) do not decouple, whereas the ghost ($c,\bar c$) and
gluon ($A=\bar A+a$) fields, as well as the heavy matter sector act as
spectators at low densities. Here we have decomposed the gauge fields
into a constant background $\bar A$ and a fluctuating part $a$.

The effective action of full QCD can then be written as
\begin{equation}\label{eq:QCDaction}
  \Gamma_k = \beta \mathcal V V[A_0] +\Delta\Gamma_k[\bar A_0, \phi]\,,
\end{equation}
where $\mathcal V$ is the spatial volume and $\beta=1/T$ the inverse
temperature.
In \eq{eq:QCDaction}, the first term denotes the QCD glue potential, encoding
the ghost-gluon dynamics in the presence of matter fields. The second term
contains the matter contribution coupled to a background gluon field
$\bar A_0$. This contribution is well-described in terms of low-energy chiral
models, such as the Nambu--Jona-Lasinio (NJL) and quark-meson (QM) models,
coupled to  Polyakov loops. In this work we make use of a Polyakov--quark-meson
(PQM) truncation~\cite{Schaefer:2007pw, Skokov:2010wb, Herbst:2010rf}
for the matter sector at low energies.
It is important to notice that the glue potential $V[A_0]$ in full QCD is
different from its Yang-Mills counterpart due to unquenching effects, see e.g.
the discussion in \cite{Herbst:2013ail, Haas:2013qwp}. The glue
potential used in effective models, on the other hand, is usually fitted to pure
Yang-Mills lattice results \cite{Pisarski:2000eq, Scavenius:2002ru,
Ratti:2005jh, Roessner:2006xn, Fukushima:2008wg, Lo:2013hla}.

To approximate unquenching effects we formulate the glue potential in terms
of the reduced temperature 
\begin{equation}\label{eq:redt}
  t=\frac{T-T_{\rm cr}}{T_{\rm cr}}\,,
\end{equation}
and write $V_{\rm YM/glue}[A_0;t]\,.$ To be more precise, there are also
two reduced temperatures, defined in terms of the critical temperatures $T_{\rm
cr}=T_{\rm cr}^{\rm YM}$ and $T_{\rm cr}= T_{\rm cr}^{\rm glue}$. One important
effect of dynamical matter fields is to lower the scale $T_{\rm cr}^{\rm glue}$
as compared to $T_{\rm cr}^{\rm YM}$, which can be used to
model the unquenching effects \cite{Schaefer:2007pw, Herbst:2010rf}.
In the present work we remedy the scale mismatch with the help of
first-principles QCD results, see \cite{Haas:2013qwp} for a detailed
discussion.
There, the FRG results for the glue potential in YM theory
\cite{Braun:2007bx, Braun:2010cy, Fister:2013bh} and QCD with two massless
quark flavours \cite{Braun:2009gm} have been compared, see also
\cite{Fischer:2013eca} for results with $2+1$ flavours.
It was found that, apart from a rescaling, the shape of the glue potential in
both theories is very similar close to $T_{\rm cr}$, see Figs.~5 and 7 in
\cite{Haas:2013qwp}. The simple linear relation 
\begin{equation}\label{eq:tYMglue}
  t_{\rm YM }(t_{\rm glue}) \approx 0.57\, t_{\rm glue}\,,
\end{equation}
is already capable of connecting the scales of both theories. 
In this manner, a potential $V_{\rm glue}[\Phi,\Phibar; t]= V_{\rm
YM}[\Phi(A_0),\Phibar(A_0); t_{\rm YM }(t)]$ is
defined,  where $\Phi,\ \Phibar$ denote the Polyakov loop and its conjugate.
Note that the relation \eq{eq:tYMglue} holds only for small and moderate
temperatures, as the slope saturates at high scales, where the perturbative
limit is reached.

In the following, \eq{eq:tYMglue} is used to account for the scale
mismatch introduced by the fit of the PQM glue potential to YM lattice
data.  The only quantity left to fix is then the critical temperature
of the glue sector, $T_{\rm cr}^{\rm glue}$. This value can in
principle also be deduced from the QCD glue potential, see
\cite{Braun:2009gm}, and yields $T_{\rm cr}^{\rm
  glue}(N_f=2)=203$~MeV.  Since the absolute scale in
\cite{Braun:2009gm} was not computed in a chiral extrapolation of the
theory with physical quark masses, we consider $T_{\rm cr}^{\rm glue}$
as an open parameter in the range
\begin{equation}
  180~\text{MeV}\lesssim T_{\rm cr}^{\rm glue} \lesssim T_{\rm cr}^{\rm YM} =
  270~\text{MeV}\,,
\end{equation}
constrained by the estimates in \cite{Schaefer:2007pw, Herbst:2010rf}.

\section{Polyakov--Quark-Meson Model}\label{sec:PQM2p1}

In the following we provide some details of the
Polyakov--quark-meson (PQM) model~\cite{Schaefer:2007pw, Skokov:2010wb,
Herbst:2010rf} and discuss the corresponding FRG flow equation at leading
order in an expansion in derivatives.

The chiral sector of this model is given by the well-known quark-meson
model~\cite{Ellwanger:1994wy, Jungnickel:1995fp, Schaefer:1997nd,  
Berges:1997eu, Schaefer:2004en}.  The integration of the gluonic
degrees of freedom results in a potential for the Polyakov-loops
($\Phi(A_0),\,\Phibar(A_0)$).  They are coupled to the matter sector via the
quark fields.

\subsection{Setup}\label{subsec:PQM}
The Euclidean Lagrangian for the PQM model reads
\begin{eqnarray} \label{eq:Lpqm}
  \mathcal L_{\rm PQM} & = & \bar{q} \left(\dr + h\,T^a (\sigma_a 
  + i \gamma_5 \pi_a ) + \mu\gamma_0\right) q \nonumber \\[1ex]
  &&  +\ \mathcal L_{m} +\ V_{\rm glue}(\Phi,\Phibar;t) \,, 
\end{eqnarray}
with a flavour-blind Yukawa coupling $h$ and the covariant derivative
$\dr(\Phi)=\gamma_\mu \partial_\mu-i\,g \gamma_0 A_0(\Phi)$ coupling
the quark fields to the Polyakov loop. In this work we assume isospin symmetry
in the light sector and use a flavour-blind chemical potential $\mu\,.$ The
mesonic Lagrangian is given by~\cite{Lenaghan:2000ey, Schaefer:2008hk,
Mitter:2013fxa}
\begin{eqnarray}\label{eq:Lmeson}
  \mathcal L_m & = & \Tr(\partial_\mu \Sigma\partial_\mu\Sigma^\dagger) +
  U(\rho_1,\tilde\rho_2) + c\,\xi \nonumber\\
  & & -\ \Tr\left[C(\Sigma+\Sigma^\dagger)\right]\,.
\end{eqnarray}
Here, the field $\Sigma$ is a complex $(3\times3)$-matrix 
\begin{equation}
  \Sigma = \Sigma_a T^a = (\sigma_a + i \pi_a)T^a\,,
\end{equation}
where $\sigma_a$ denotes the scalar and $\pi_a$ the pseudo-scalar meson nonets
and the Hermitian generators of the flavour $U(3)$ symmetry are defined via the
Gell-Mann matrices as $T^a=\lambda^a/2\,.$ It is advantageous to rotate into the
non-strange--strange basis via
\begin{equation}
  \begin{pmatrix}
    \sigma_x\\ 
    \sigma_y
  \end{pmatrix} = \frac{1}{\sqrt{3}} 
  \begin{pmatrix}
    \sqrt{2} & 1 \\
    1 & -\sqrt{2}
  \end{pmatrix}
  \begin{pmatrix}
    \sigma_0\\
    \sigma_8
  \end{pmatrix}\,,
\end{equation}
with $\sigma_x$ the non-strange and $\sigma_y$ the strange condensate.
Then, the explicit symmetry breaking term consistent with isospin symmetry 
takes the simple form
$\Tr\left[C(\Sigma+\Sigma^\dagger)\right]\to c_x\sigma_x + c_y\sigma_y$\,, with
$c_x, c_y$ governing the bare light and strange quark masses, respectively.

The meson potential $U$ can be expressed via the chiral invariants $\rho_i =
\Tr\left[(\Sigma\Sigma^\dagger)^i\right]$, $i=1,\,\dots,\,N_f$ 
\cite{Jungnickel1996b}. In the ($2+1$)-flavour approximation, where
$\sigma_3=0$, $\rho_3$ can be expressed in terms of the other invariants and we
use the set $\lbrace \rho_1,\,\tilde\rho_2,\,\xi\rbrace$ with 
\begin{eqnarray}
  \rho_1 & = & \frac{1}{2}\left(\sigma_x^2 + \sigma_y^2\right)\,,\nonumber\\
  \tilde\rho_2 & = & \rho_2 - \frac{1}{3}\rho_1^2 = \frac{1}{24}\left(\sigma_x^2
  - 2 \sigma_y^2 \right)^2\,,\nonumber\\
  \xi & = & \det(\Sigma) +\det(\Sigma^\dagger)=\frac{\sigma_x^2\sigma_y}{2 \sqrt{2}}\,.
\end{eqnarray}
Here, $\xi$ represents the 't\,Hooft determinant~\cite{Hooft:1976fv,
Hooft:1976up}, rewritten in the mesonic language~\cite{Kobayashi:1971qz,
Hooft:1986nc}, and as such implements the chiral $U_A(1)$ anomaly.
The strength of its coupling, $c$, determines the mass splitting between the
$\eta$, $\eta'$ and pions, see e.g. \cite{Mitter:2013fxa, Schaefer:2008hk,
Schaefer:2013isa} for a detailed discussion.

Furthermore, the quasi-particle energies of the quarks and mesons are given
by $E_i = \sqrt{k^2+m_i^2}$, $i~\in~\lbrace l,\, s,\, j \rbrace$ with $j \in
\lbrace \sigma,\, a_0,\, \kappa,\, f_0,\, \pi,\, K,\, \eta,\, \eta' \rbrace$.
The masses themselves are defined as
\begin{eqnarray}
  \label{eq:quarkmasses}
  m_l & = & h\frac{\langle\sigma_x\rangle}{2}\,,\nonumber\\
  m_s & = & h\frac{\langle\sigma_y\rangle}{\sqrt{2}}\,,
\end{eqnarray}
for the light and strange quarks, respectively, and
\begin{equation}\label{eq:mesonmasses}
  \lbrace m_j^2\rbrace = \text{eig}\left\lbrace H_\Sigma \left(U(\rho_1,
  \tilde\rho_2) + c\xi\right) \right\rbrace
\end{equation}
for the mesons. Here, $H_\Sigma(.)$ denotes the Hessian w.r.t. $\Sigma$ and
$\text{eig}\lbrace .\rbrace$ denotes the set of eigenvalues of the given
operator.  For further details on this model we refer the reader to
\cite{Lenaghan:2000ey, Schaefer:2008hk, Mitter:2013fxa}.
The  two-flavour case considered in \sec{subsec:TD2} is obtained by omitting the
strange quark sector as well as all mesons except the sigma and pions.

What is now left is to specify the glue potential $V_{\rm glue}$. We have argued
above that we can use the  YM-based parametrisations $\mathcal{U}$ of the glue
potential and modify the scale according to QCD FRG results, \eq{eq:tYMglue}.
Several parametrisations of the Polyakov-loop potential have been put forward
in the recent years \cite{Pisarski:2000eq, Scavenius:2002ru, Ratti:2005jh,
Roessner:2006xn, Fukushima:2008wg, Fukushima:2012qa, Lo:2013hla}.
In the main text we only show results for a polynomial version, introduced
in~\cite{Pisarski:2000eq, Ratti:2005jh}
\begin{eqnarray} \label{eq:upoly}
  \frac{\mathcal U_{\rm poly}(\Phi,\Phibar;t)}{T^{4}} & = &
  -\frac{b_2(t)}{2}\Phi\Phibar 
   -\ \frac{b_3}{6}\left(\Phi^{3}+\Phibar^{3}\right) \nonumber\\
  && +\ \frac{b_4}{4} \left(\Phi\Phibar\right)^{2}\,.
\end{eqnarray}
The temperature-dependent coefficient, expressed in terms of the reduced
temperature, is given by
\begin{equation}\label{eq:upolypara}
  b_2(t) =  a_0 + \frac{a_1 }{1+t} + \frac{a_2}{(1+t)^2}
+\frac{a_3}{(1+t)^3}\,. 
\end{equation}
The parameters $a_i,\, b_i$ of Eqs.~(\ref{eq:upoly}) and (\ref{eq:upolypara})
have been determined in~\cite{Ratti:2005jh} by a fit to pure Yang-Mills lattice
results to be
\begin{equation}
  a_0 = 6.75\,,\  a_1= -1.95\,, \ a_2 = 2.625\,,\ a_3 = -7.44
\end{equation}
and
\begin{equation}
  b_3=0.75\,,\quad b_4 = 7.5\,.
\end{equation}
We use the lattice result for the pressure to fix the open parameter $T_{\rm
cr}^{\rm glue}=210$~MeV in $\mathcal{U}(\Phi,\Phibar;t)$. A discussion of the
dependence of our results on this choice and on the parametrisation of
$\mathcal U$ can be found in the appendix.

\subsection{Fluctuations in the PQM model}\label{subsec:PQMflow}

In the present work we go beyond the mean-field approximation used
in~\cite{Haas:2013qwp} and apply the FRG to include quantum and
thermal fluctuations of the PQM model. This provides us with a more
realistic description of the chiral and deconfinement phase
transitions.  In fact it has been shown previously, see
e.g.~\cite{Schaefer:2004en}, that fluctuations smear out the phase
transition, yielding smoother transitions that are in better agreement
with lattice results.

The flow equation for the two-flavour PQM model has been derived previously
in~\cite{Skokov:2010wb,Herbst:2010rf} while the flow of the $(2+1)$-flavour
quark-meson model is discussed in depth in~\cite{Mitter:2013fxa}. It is then
straight-forward to deduce the flow equation of the full $(2+1)$-flavour PQM
model 
\begin{eqnarray}\label{eq:PQM2+1flow}
   & & \partial_t \Omega_k = \frac{k^5}{12\pi^2}
  \left\lbrace \sum_{i=1}^{2 N_f^2}\frac{1}{E_i}\coth\left(\frac{E_i}{2T}\right)
   \right.\\[1ex]
  && \quad \left. -\frac{8N_c}{E_{l}}\left[1 - N_{l}( T,\mu;\Phi,\Phibar)
    -  N_{\bar l} ( T,\mu; \Phi,\Phibar)\right] \right.\nonumber\\[1ex]
  && \quad \left. -\frac{4N_c}{E_{s}}\left[1 - N_{s}( T,\mu; \Phi,\Phibar)
    - N_{\bar s} ( T,\mu; \Phi,\Phibar)\right] \right\rbrace\,.\nonumber
\end{eqnarray}
The Polyakov-loop extended quark/anti-quark occupation numbers are given by
\begin{eqnarray}
  N_{q}(T,\mu;\Phi,\Phibar) = \hspace{148pt} &&\\ 
  \dfrac{1+2\Phibar e^{(E_q-\mu)/T}+\Phi e^{2(E_q-\mu)/T}}{1+3\Phibar
  e^{(E_q-\mu)/T}+3\Phi e^{2(E_q-\mu)/T}+e^{3(E_q-\mu)/T}}\,,&&\nonumber
\end{eqnarray}
and $N_{\bar q}(T,\mu;\Phi,\Phibar) \equiv N_q(T,-\mu;\Phibar,\Phi)$
for $q=l,s$.

Note that we restrict ourselves to leading order in a derivative
expansion and neglect the running of any couplings involving quark interactions.
The RG running of the mesonic couplings, on the other hand, is encoded in the
scale-dependent effective potential $\Omega_k$.

In order to solve the flow \eq{eq:PQM2+1flow}, we have to specify an
initial potential at the cutoff scale $\Lambda$. In this work we have
chosen $\Lambda=1$~GeV, in accordance with our interpretation of the
quark-meson model as a low-energy effective description.  We keep only
renormalisable terms in the mesonic potential at the cutoff scale and
restrict ourselves to only two chiral invariants $\rho_1,
\tilde\rho_2$, cf. the discussion in~\cite{Mitter:2013fxa}
\begin{equation}
  U_\Lambda(\rho_1,\tilde\rho_2) = a_{10}\rho_1 + a_{01}\tilde\rho_2 +
  \frac{a_{20}}{2}\rho_1^2\,.
\end{equation}
The parameters are fixed to $a_{10}=(972.63~\text{ MeV})^2,\, a_{01}=50,\,
a_{20}=2.5$ which, together with the choices $h~=~6.5$ for the Yukawa coupling,
$c=-4807.84$~MeV for the 't\,Hooft--determinant coupling and explicit breaking
strengths $c_x=(120.73~\text{MeV})^3$ and $c_y=(336.41~\text{MeV})^3$,
reproduces the physical spectrum as well as the pion and kaon decay constants in
the vacuum \cite{Beringer:1900zz}. In particular, we have chosen a sigma-meson 
mass of $m_\sigma = 400$~MeV. In Appendix~\ref{app:msigma} we discuss the
dependence of our results on this choice.

At temperatures $2\pi T\gtrsim \Lambda\,,$ thermal fluctuations become important
also at scales above the cutoff $k>\Lambda$.
These thermal fluctuations are, however, not taken into account in the solution
to the flow \eq{eq:PQM2+1flow} with finite cutoff $\Lambda$. Therefore, the
initial potential $\Omega_\Lambda$ is not fully independent of temperature,
which is quantitatively important in the region  $2\pi T\gtrsim \Lambda$. 
On the other hand, this temperature dependence of the initial potential
$\Omega_\Lambda$ is also governed by the flow \eq{eq:PQM2+1flow} and can be
obtained by integrating the vacuum flow from the cutoff $\Lambda$ up to a scale
$\bar\Lambda\gg 2\pi T$ and subsequently integrating the finite temperature
flow down to $\Lambda$ again
\begin{equation}
  \Delta\Omega_{\Lambda}(T,\mu) = \int_{\Lambda}^{\bar\Lambda} \frac{dk}{k}
  \left(\partial_t \Omega_k(T,\mu) - \partial_t\Omega_k(0,0) \right)\,.
\end{equation}
This procedure is equivalent to a change of the initial scale from $\Lambda$ to
$\bar\Lambda$, while keeping the infrared physics fixed, i.e. a change in the
renormalisation scale. However, as we expect mesonic fluctuations to be less
important at scales $k>\Lambda=1$ GeV, we approximate the difference 
by the purely fermionic contribution to \eq{eq:PQM2+1flow}. Since the fermionic
contribution to the flow is independent of $\Omega_k$, the approximate
temperature dependence of $\Omega_\Lambda$ is given by the simple integral
\cite{Braun:2003ii, Herbst:2010rf} 
\begin{eqnarray}                                          
  \Delta\Omega_\Lambda & = & \left.\int^\infty_\Lambda dk\frac{k^4}{12\pi^2}
    \right\lbrace \\
    & &  \quad \quad \frac{8 N_c}{E_l}\left[N_{l}( T,\mu;\Phi,\Phibar) +
      N_{\bar l} (T,\mu; \Phi,\Phibar)\right]\nonumber\\[1ex]
  && \quad \left. +\ \frac{4N_c}{E_{s}}\left[N_{s}( T,\mu; \Phi,\Phibar)
    + N_{\bar s} ( T,\mu; \Phi,\Phibar)\right] \right\rbrace\,.\nonumber
\end{eqnarray}
Here, we have chosen $\bar\Lambda=\infty$, since the fermionic
difference flow is finite. Finally we obtain
\begin{eqnarray}\label{eq:UVpot}
  \Omega_\Lambda(T,\mu;\sigma_x,\sigma_y,\Phi,\Phibar) & = &
    U_\Lambda(\rho_1,\tilde\rho_2) +\, \mathcal{U}(\Phi,\Phibar;t)
    \nonumber\\[1ex]
     &  &\hspace{-24pt} +\ \Delta\Omega_\Lambda(T, \mu; \sigma_x, \sigma_y,
      \Phi,\Phibar)\, ,
\end{eqnarray}
for the initial potential at the cutoff scale $\Lambda$,
including fermionic temperature corrections.
\begin{figure}
  \includegraphics[width=.49\textwidth]{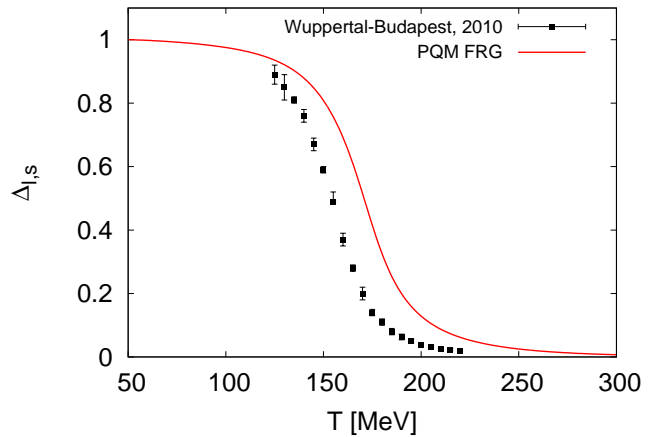}
  \caption{Temperature dependence of the subtracted chiral condensate: the FRG
  curve is compared to the lattice result by the Wuppertal-Budapest
  collaboration \cite{Borsanyi:2010bp}.}
  \label{fig:condensateT}
\end{figure}
%

\section{Results}\label{sec:results}

\subsection{QCD Crossover}\label{subsec:crossover}
\begin{figure*}
  \includegraphics[width=.49\textwidth]{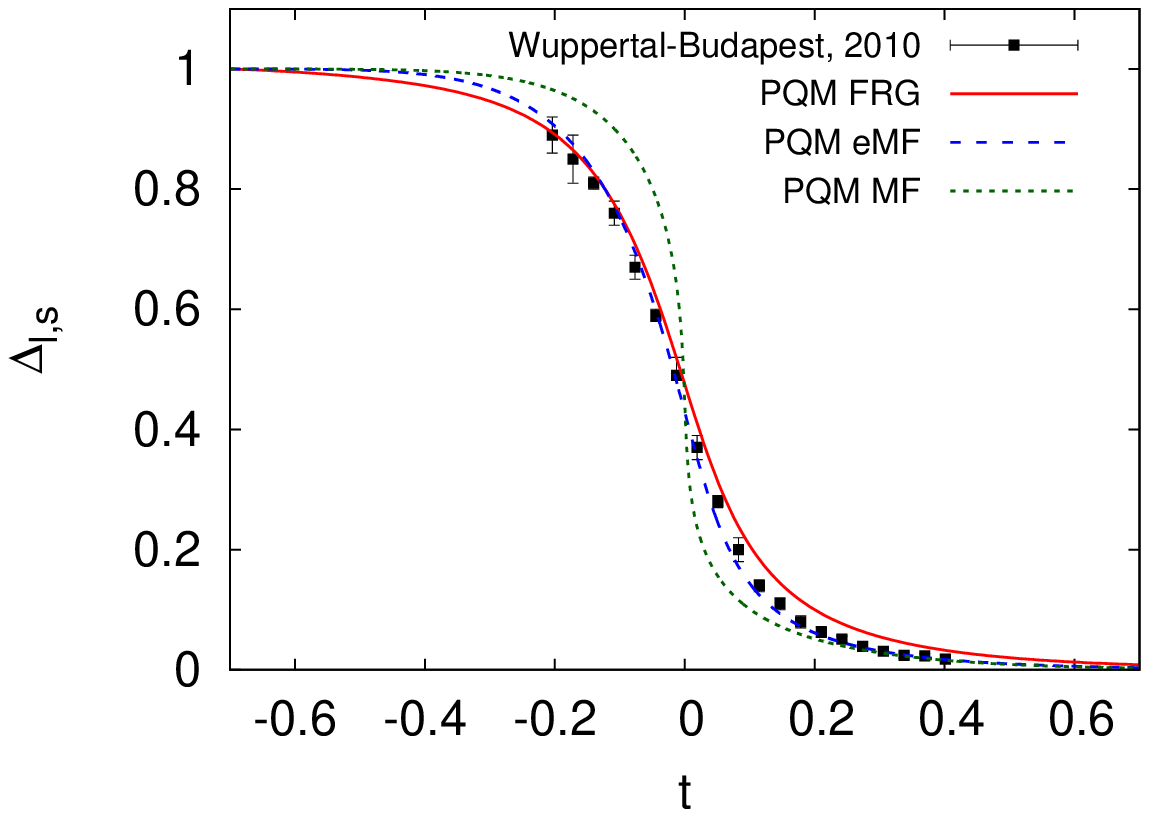}
  \includegraphics[width=.49\textwidth]{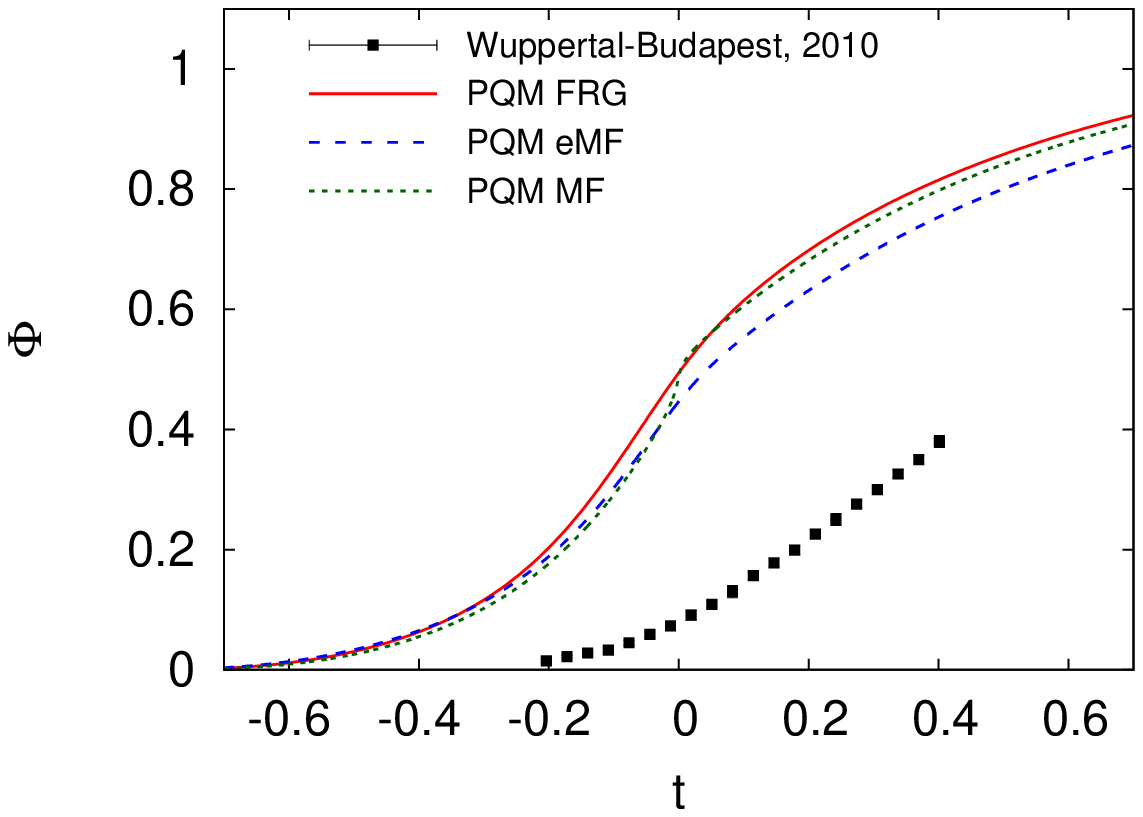}
  \caption{Temperature dependence of the subtracted chiral condensate (left)
  and  Polyakov loop (right). The FRG result is compared to
  the lattice result of the Wuppertal-Budapest collaboration,
  \cite{Borsanyi:2010bp}, as well as to the mean-field result. See text for
  comments on the Polyakov loop in continuum approaches.}
  \label{fig:order_params}
\end{figure*}
From the solution to the flow \eq{eq:PQM2+1flow} we can determine the phase
structure and thermodynamics of the PQM model. For the time being, we
restrict ourselves to vanishing chemical potential. This has the advantage that
in this limit the Polyakov loop and its conjugate coincide, $\Phibar = \Phi$.
Hence, the numerical effort to solve the equations of motion (EoM)
\begin{equation}\label{eq:EoM}
  \left.\dfrac{\partial\Omega_{k\to0}}{\partial\sigma_x}\right|_{\chi_0}
  = \left.\dfrac{\partial\Omega_{k\to0}}{\partial\sigma_y}\right|_{\chi_0}
  = \left.\dfrac{\partial\Omega_{k\to0}}{\partial\Phi}\right|_{\chi_0}= 0\,,
\end{equation}
which determine the order parameters $\chi_0 = \left(\sigma_x,  \sigma_y,
\Phi\right)$ for given temperature and chemical potential, is drastically
reduced. A discussion of the numerical method used to solve this
multi-dimensional system of partial differential equations can be found
in~\cite{Strodthoff:2011tz, Mitter:2013fxa}.
\begin{figure*}
  \includegraphics[width=.49\textwidth]{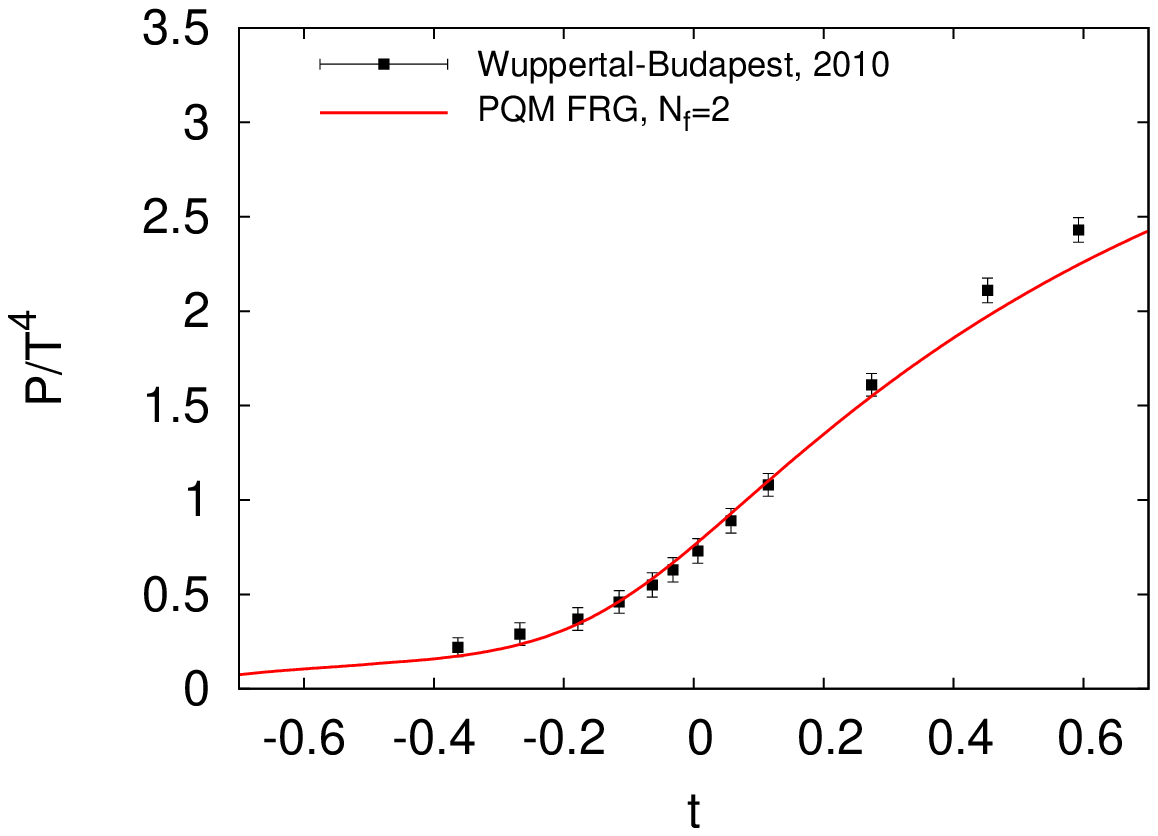}
  \includegraphics[width=.49\textwidth]{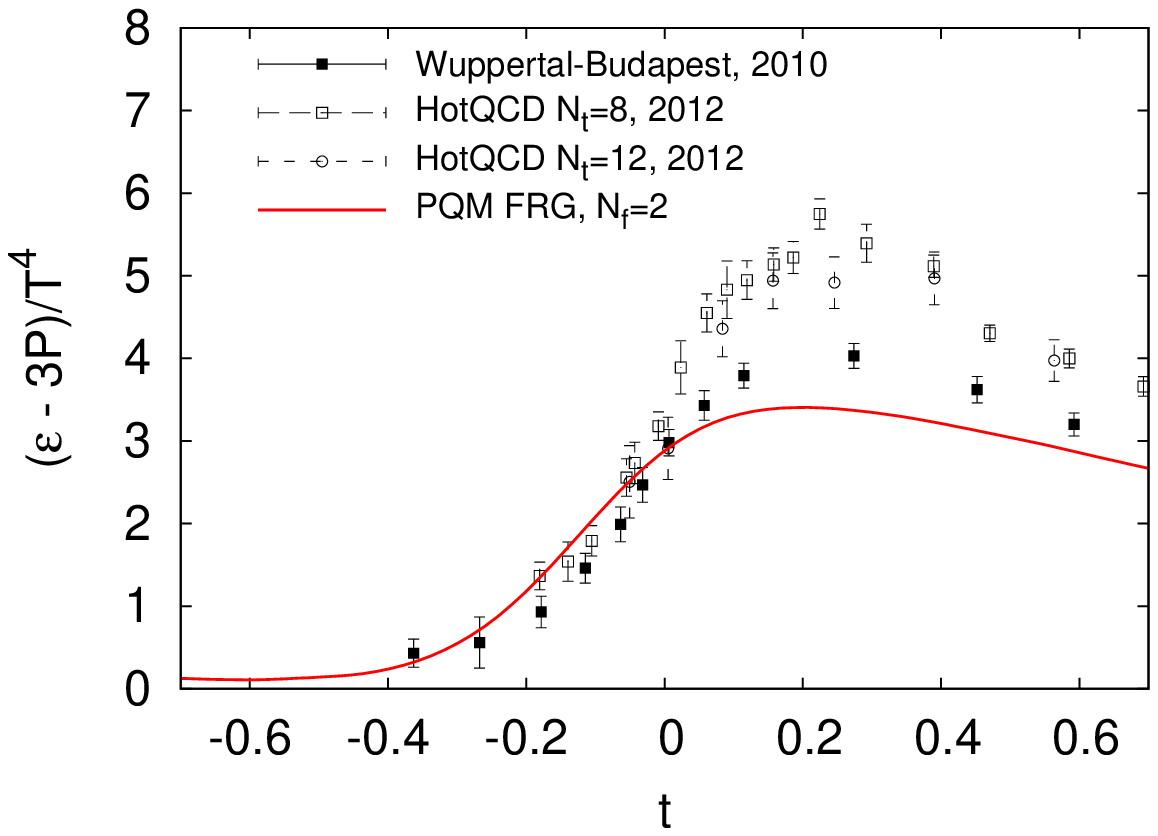}
  \caption{FRG result for the pressure (left) and interaction measure (right)
    for two quark flavours, compared to the $(2+1)$-flavour lattice results
    \cite{Bazavov:2012bp, Borsanyi:2010cj}. See text for details.}
  \label{fig:Nf2_pdelta}
\end{figure*}

In \fig{fig:condensateT} our result for the subtracted chiral
condensate,
\begin{equation}
 \Delta_{l,s}  = \frac{\left(\sx-\frac{c_{x}}{c_{y}}\sy\right)_T}{ \left(
 \sx-\frac{c_{x}}{c_{y}} \sy\right)_{T=0}}
\end{equation} 
is shown in comparison with the lattice result by the
Wuppertal-Budapest collaboration \cite{Borsanyi:2010bp}.  Due to the
finite quark masses, we find a smooth crossover and there is no exact definition
of the transition temperature. Nevertheless, it is customary to associate a
transition temperature with the peak position of the temperature derivative of
the order parameter, $d\Delta_{l,s}/dT$. Using this definition, we
obtain $T_\chi=172$~MeV for the chiral crossover temperature, and similarly
$T_d=163$~MeV for the Polyakov-loop related transition via $d\Phi/dT$. Both
values agree roughly with the transition region on the lattice $(147-165)$~MeV
\cite{Borsanyi:2010bp} and the pseudocritical temperatures
$T_\chi=157\pm{3}$~MeV \cite{Borsanyi:2010bp} and $T_\chi=154\pm{9}$~MeV
\cite{Bazavov:2011nk}.

Note that, apart from a shift along the temperature axis,the slope
of our FRG result for the subtracted condensate coincides with the lattice one,
cf. Figs.~\ref{fig:condensateT} and \ref{fig:order_params}. This indicates that
the relative strength of the relevant dynamics is included properly. However,
there is a difference in the absolute scale, $T_\chi$, in our calculation and
the lattice. This is to some extent related to our choice of the sigma meson
mass.  From experiment it is known that the $\sigma$ ($f_0(500)$) is a broad
resonance, $(400-550) - i(200-350)$~MeV \cite{pdg:2012}. It has been shown
previously, see e.g.~\cite{Schaefer:2008hk}, that a lower sigma mass results
in a lower chiral transition temperature, while the slope of the condensate is
only changed marginally.
For the curve shown in \fig{fig:condensateT} the value $m_\sigma=400~$MeV has
been used. We have checked that for lower values, entailing stronger mesonic
fluctuations, the result is shifted even closer to the lattice one. However, the
value of $m_\sigma=400$~MeV is already at the lower experimental boundary, hence
we refrain from using a lower mass in the following. The impact of this mass
parameter on thermodynamics is also discussed in Appendix~\ref{app:msigma}. 

The axial anomaly similarly influences the transition. It has been demonstrated
in \cite{Mitter:2013fxa} that the transition temperature is reduced for
vanishing anomaly coupling, $c=0$. In fact, with our choice of
$m_\sigma=400~$MeV and $c=0$, the resulting condensate lies almost exactly on
top of the lattice points. Note that in this case the $\eta'$ meson
would be an additional pseudo-Goldstone boson, again leading to enhanced
fluctuations. However, this choice is unphysical and results in, e.g., a too
high pressure at low temperatures. The use of a temperature-dependent anomaly
coupling, $c(T)$, is expected to resolve this issue. In summary, we
have found that for a correct description of the absolute scale, further mesonic
fluctuations need to be included. Within our FRG treatment this would correspond
to higher mesonic operators in our cutoff potential, $\Omega_\Lambda$. In full
QCD such contributions are dynamically generated at higher scales, but we have
omitted them in the present work since we restrict our cutoff action to contain
renormalisable operators only.

We conclude that while the absolute scale, $T_\chi$, differs from the lattice one
by about $10\%$ in our FRG calculation, the relative strength of the
relevant dynamics of the transition is captured well. This is due to the
inclusion of unquenching effects as well as matter fluctuations. We postpone the
improvement of our scale-setting procedure to future work and concentrate on the
discussion of the dynamics of the transition in the following. To this end, all
results are expressed in terms of the reduced temperature $t=(T-T_\chi)/T_\chi$.
This choice allows us to compare the overall shape - and thereby the proper
inclusion of the relevant dynamics - of the observables, while a mismatch of the
critical temperatures is scaled out. 

{\fig{fig:order_params} shows the subtracted chiral condensate (left)
as well as the Polyakov loop (right) in terms of the reduced temperature. As
argued above, we observe excellent agreement between the FRG (solid line) and
lattice (symbols) result for the chiral condensate, especially at temperatures
below $T_\chi$. In turn, for temperatures above the transition the present model
overestimates the importance of mesonic fluctuations, and the FRG result for the
order parameter is above the lattice result. The use of dynamical hadronisation,
\cite{Gies:2001nw,  Gies:2002hq, Pawlowski:2005xe, Floerchinger:2009uf}, should
compensate this effect.

For comparison, in \fig{fig:order_params} we also show results from a PQM 
mean-field calculation without (``MF'', dotted line) and with (``eMF``, dashed
line) the fermionic vacuum loop contribution \cite{Skokov:2010sf,
Andersen:2011pr, Gupta:2011ez, Schaefer:2011ex}.
Note that we fix the remaining parameters of the model, $m_\sigma$ and $T_{\rm
cr}^{\rm glue}$, by comparing the pressure to the lattice pressure, cf.
discussion in Appendix \ref{app:Tcr}. In this manner, the effect of
fluctuations is partially included in the model parameters. This
results in different parameter values for the mean-field and FRG calculations.
We use $m_\sigma=500$~MeV and $T_{\rm  cr}^{\rm glue}=210$~MeV for the standard
mean-field calculation, as discussed in \cite{Haas:2013qwp, Stiele:2013gra} and
$m_\sigma=400$~MeV, $T_{\rm cr}^{\rm glue}=260$~MeV for the extended mean-field
calculation. However, it is clear from, e.g., \fig{fig:order_params}, that a
modification of the parameters is not sufficient to describe the full dynamics
of the transition. The inclusion of fluctuations, as done in our FRG setup, is
crucial to reproduce the slope of the order parameter as well as thermodynamic
observables correctly.

In fact, the pseudocritical temperature of the standard mean-field approximation
is closer to the lattice one than our FRG result, $T_\chi=T_d=158$~MeV. However,
this approach neglects mesonic fluctuations, and the transition comes out too
steep, see \fig{fig:order_params}. Including the fermionic vacuum fluctuations
yields too high pseudocritical temperatures, $T_\chi=181$~MeV and $T_d=173$~MeV.
Compared to the standard MF result, on the other hand, the slope of the
condensate is reduced.

A word of caution needs to be added concerning the Polyakov loop,
\fig{fig:order_params} (right). 
It is well-known that the definitions of this quantity used on the lattice,
$\langle\Phi\rangle$, and the present continuum formulation, $\Phi[\langle
A_0\rangle]$, differ and a direct comparison is not possible. In view of this,
we do not expect agreement of these two observables. However, it can be shown
that the continuum definition serves as an upper bound for the lattice one,
$\Phi[\langle A_0\rangle]\geq~\langle\Phi\rangle\,,$ up to renormalisation
issues, cf. \cite{Braun:2007bx, Marhauser:2008fz}.
Hence, an approximate coincidence of the respective crossover regions is
still anticipated. Indeed, we find that our transition temperature, defined by
the inflection point of the Polyakov loop, roughly agrees with the transition
region found on the lattice.
\begin{figure*}
  \includegraphics[width=.49\textwidth]
  {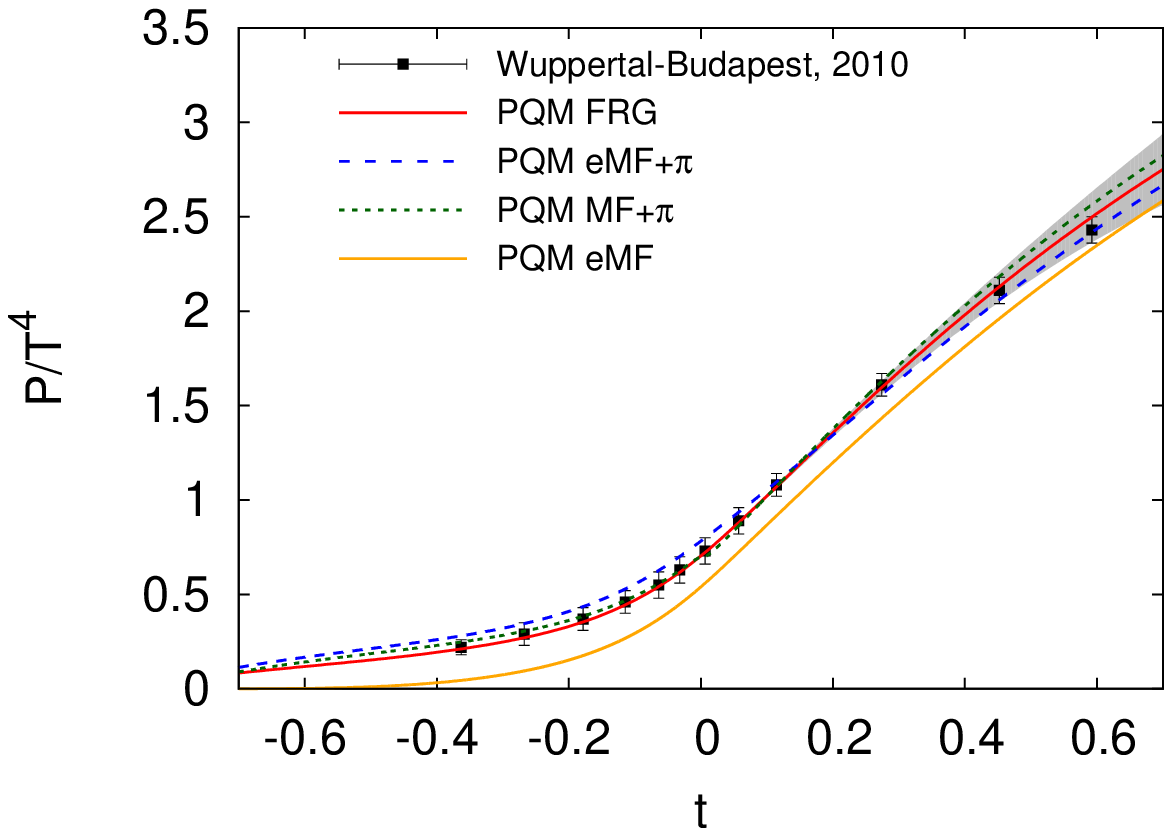}
  \includegraphics[width=.49\textwidth]
  {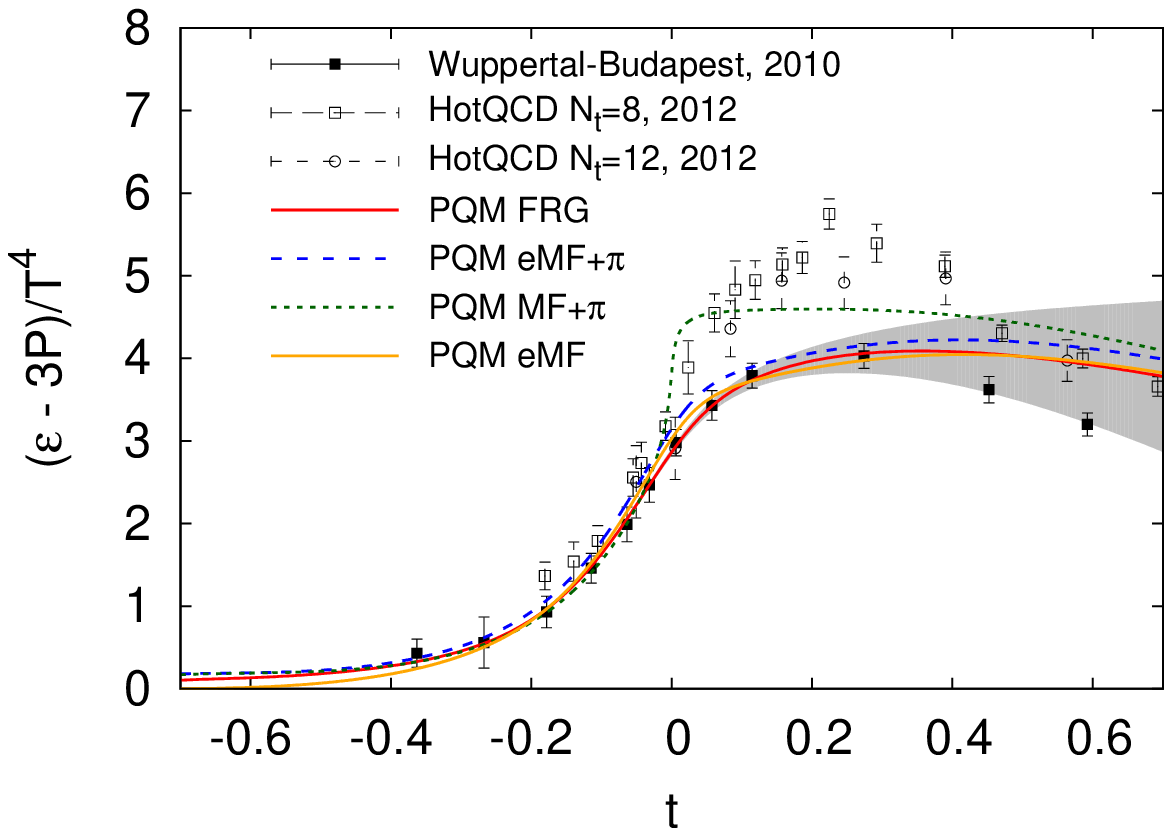}
  \caption{$(2+1)$-flavour FRG results for the pressure (left) and interaction
  measure (right) compared to the lattice, \cite{Bazavov:2012bp,
  Borsanyi:2010cj}, and  mean-field results. See text for
  details.}
  \label{fig:Nf2p1_pdelta_poly}
\end{figure*}
%

\subsection{Thermodynamics: $N_f=2$}\label{subsec:TD2}
Within the FRG framework, the full quantum effective potential is defined by the
effective average potential $\Omega_k$ in the infrared, evaluated on the
solution of the EoM,
\begin{equation}
  \Omega(T,\mu) = \Omega_{k\to0}(T,\mu)|_{\chi_0}\,.
\end{equation}
The pressure of the system is then given by the negative of the effective
potential, normalised in the vacuum
\begin{equation}
  P = -\Omega(T,\mu) + \Omega(0,0)\,,
\end{equation}
and serves as a thermodynamic potential, from which we can deduce other bulk
thermodynamic quantities in the standard way. In particular, we are interested
in the free energy density
\begin{equation}
  \epsilon = -P + T s + \sum_f \mu_f n_f \,,
\end{equation}
with the entropy density $s=\partial P/\partial T$ and the quark number
densities $n_f = \partial P/\partial \mu_f $ for $f=u,d,s$\,. Moreover,
we consider the interaction measure
\begin{equation}
  \Delta = \epsilon - 3 P\,,
\end{equation}
which quantifies the deviation from the equation of state of an ideal
gas, $\epsilon = 3 P$.

We compare these quantities to results of the HotQCD collaboration,
\cite{Bazavov:2012bp}, using the HISQ action and temporal lattice extents of
$N_\tau = 8, 12$ as well as to the continuum extrapolated results of the
Wuppertal-Budapest collaboration \cite{Borsanyi:2010cj}.

We start our discussion of thermodynamic quantities by studying the
two-flavour case. The flow equation of the two-flavour PQM model has previously
been discussed in \cite{Skokov:2010wb, Herbst:2010rf}. Here, we
use the parameter set given in \cite{Herbst:2013ail}, with $m_\sigma=540$~MeV.
To our knowledge, there are no recent two-flavour lattice results for
thermodynamics available. This entails that we cannot fix the remaining
parameter, $T_{\rm cr}^{\rm glue}$, to, e.g., the lattice pressure. Instead we
have chosen the same value as for $2+1$ flavours, $T_{\rm cr}^{\rm glue}=
210$~MeV, see also our discussion in \sec{subsec:TD2+1} below. This
value is close to the phenomenological HTL estimate, $T_{\rm
cr}^{\rm glue}(N_f=2)= 208$~MeV put forward in~\cite{Schaefer:2007pw,
Herbst:2010rf} and the FRG estimate, $T_{\rm cr}^{\rm glue}=203$~MeV
of~\cite{Braun:2009gm}. Furthermore, this choice results in almost
degenerate chiral and Polyakov-loop critical temperatures. Having fixed all
parameters, the results of the present section serve as a prediction for the
thermodynamics of two-flavour QCD. 

In \fig{fig:Nf2_pdelta} we show the pressure (left) and interaction measure
(right), both normalised by $T^4$. For comparison we also show the
lattice results for $2+1$ flavours. Despite the fact that we only consider two
quark flavours here, the overall agreement is rather good. At low temperatures,
the lightest mesonic degrees of freedom, the pions, are expected to dominate the
pressure. These are already included in the two-flavour model. At high
temperatures, on the other hand, the two-flavour FRG result underestimates the
$(2+1)$-flavour lattice value. This is expected due to the additional third
quark species contributing to the lattice pressure at high $T$.

The right panel of  \fig{fig:Nf2_pdelta} shows the interaction measure.
Deviations from the lattice are more pronounced in this quantity due to the
presence of derivatives in its definition. Of course we do not expect perfect
agreement between our two-flavour computation and the $N_f=2+1$ lattice result.
However, the strongest modifications are expected around the phase transition,
where there are more light degrees of freedom contributing to the thermodynamics
for $2+1$ flavours.
In the low and high temperature regimes the quarks and mesons are heavy,
respectively, and hence contribute less to the thermodynamic observables. This
explains the surprisingly good agreement between the two and three flavour
results.
In fact, we find reasonable agreement with the lattice data below the phase
transition, $t\leq 0$. While the peak height is underestimated, the increase in
$\Delta/T^4$ around $T_c$ is similar to the lattice. Above $T_c$, the
two-flavour curve lies below the lattice result.

\subsection{Thermodynamics: $N_f= 2+1$}\label{subsec:TD2+1}
Next, we turn to the (2+1)-flavour model. Here, we can directly
compare to the available (2+1)-flavour lattice results and fix our open
parameter, $T_{\rm cr}^{\rm glue}\,$ by comparison of the pressure.
\fig{fig:Nf2p1_pdelta_poly} (left, solid line) then shows our result for the
pressure, normalised by $T^4$, which agrees very well with the lattice
result in the continuum limit. 
Near $t=0$ this is a consequence of our choice of
$T_{\rm cr}^{\rm glue}$. The nice agreement with lattice data away from
$T_\chi$, on the other hand, indicates that we have included all relevant
degrees of freedom, especially below the transition temperature.
The grey band gives an error estimate of our FRG result, which is obtained
from the change of the threshold functions with respect to the temperature, at
vanishing mass at the ultraviolet cutoff $\Lambda$. This results in
\begin{eqnarray}
 P\pm\Delta P(\Lambda,T) & = & P\left(1\pm\frac{2}{e^{\Lambda/T}-1}\right)\,.
\end{eqnarray}
The propagation of uncertainty in the interaction measure as a derived
quantity has been taken into account via
\begin{eqnarray}
 \frac{d\left(P\pm\Delta P(\Lambda,T)\right)}{dT} & = & \frac{dP}{dT}
  \pm\frac{d\left(P\frac{2}{e^{\Lambda/T}-1}\right)}{dT}\,.
\end{eqnarray}

Also shown in \fig{fig:Nf2p1_pdelta_poly} are the mean-field results.
To achieve a better description of the thermodynamics at low $T$, we have
augmented the MF
and eMF results by the contribution of a thermal pion gas, where the pion
in-medium mass is determined by the mean-field potential.
Our two flavour FRG calculation, see \sec{subsec:TD2}, confirms that  these are
the relevant degrees of freedom below the phase transition. To highlight the
impact of this pion contribution, we also show results for the pure eMF
calculation in \fig{fig:Nf2p1_pdelta_poly} (yellow, solid line).
Strictly speaking, this contribution picks up a field-dependence via the
in-medium pion mass, which would modify the equations of motion. Here, however,
we consider it as a correction to the thermodynamic potential only, and hence
neglect its backcoupling on the equations of motion. For consistency, we also
neglect all terms containing field derivatives, $\pd{P_\pi}/\pd\phi_i$, in
higher thermodynamic observables. 

While the pressure of the mean-field approximation including pions (dotted line)
lies above the FRG and lattice ones, the inclusion of the vacuum term (dashed
line) results in an additional increase in $P$ at low $t$ and a decrease at high
$t$.
As expected, the omission of the pion contribution in the eMF calculation
(solid, yellow line) yields a pressure that is too low, especially below the
phase transition, where pions are expected to dominate the thermodynamics.

The interaction measure is displayed in the right panel of
\fig{fig:Nf2p1_pdelta_poly}. 
Similarly to the other observables, also the interaction measure is too steep
within the standard mean-field approximation. Including the vacuum term, the
transition is smoothened out and already agrees quite well with the
Wuppertal-Budapest results.
Turning to our FRG result (red, solid curve), we find remarkably good
agreement with the continuum extrapolated lattice result from the
Wuppertal-Budapest collaboration. There is a stronger deviation from the HotQCD
data, but we attribute this to the lacking continuum limit of their data.
In fact it is observed that the peak height of $\Delta/T^4$ goes down as the
continuum is approached, cf. \cite{Borsanyi:2010cj, Bazavov:2012bp}.
Although not shown explicitly, we want to stress that the drastic
reduction of the peak height in the interaction measure towards the lattice
results is due the inclusion of $t_{\rm YM}(t_{\rm glue})$ in both the
FRG and mean-field approaches, see also \cite{Haas:2013qwp, Stiele:2013gra}.

In comparison to our two-flavour result, we see that the inclusion
of the heavier strange quark and especially the full scalar and pseudoscalar
meson nonets increases the peak height of the interaction measure and puts our
curve right on top of the lattice result.
At high temperatures we find that $\Delta/T^4$ decreases too slowly in our
calculation. However, this is the region where our scale matching procedure,
\eq{eq:tYMglue}, ceases to be valid and corrections are expected. 
\begin{figure*}
  \includegraphics[width=.49\textwidth]{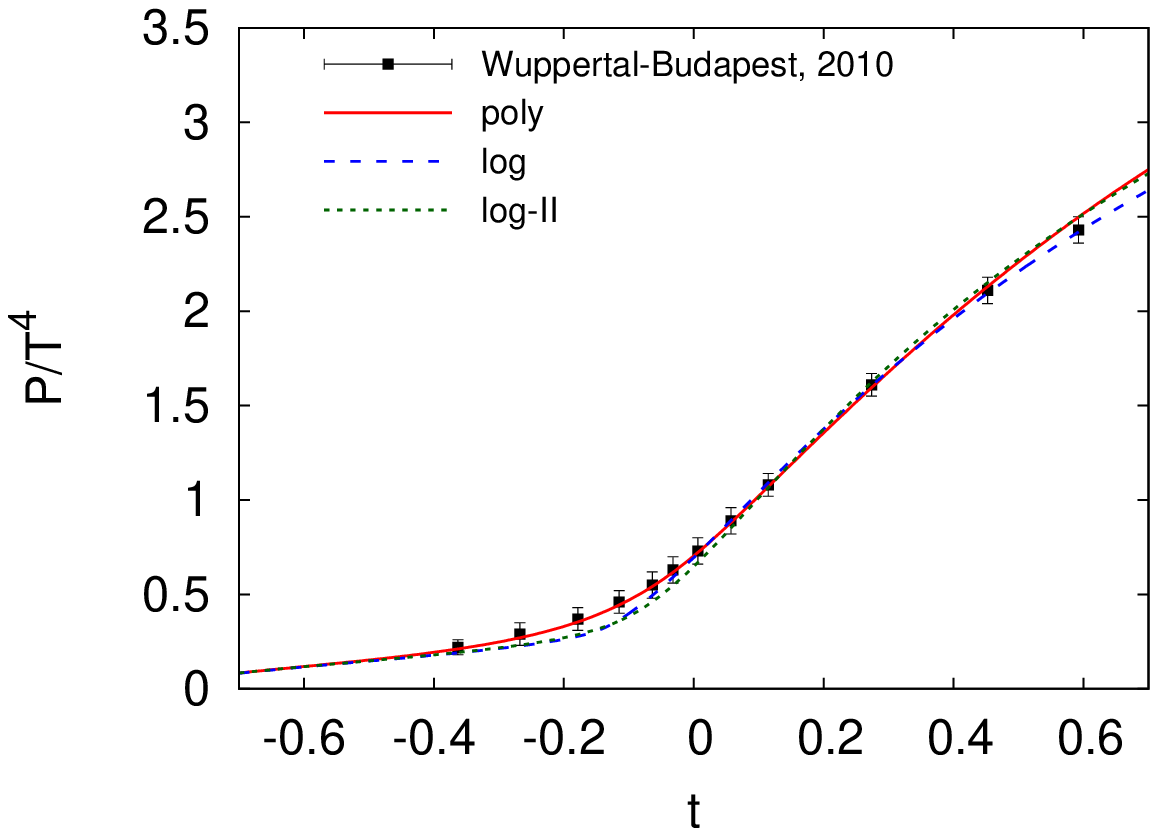}
  \includegraphics[width=.49\textwidth]{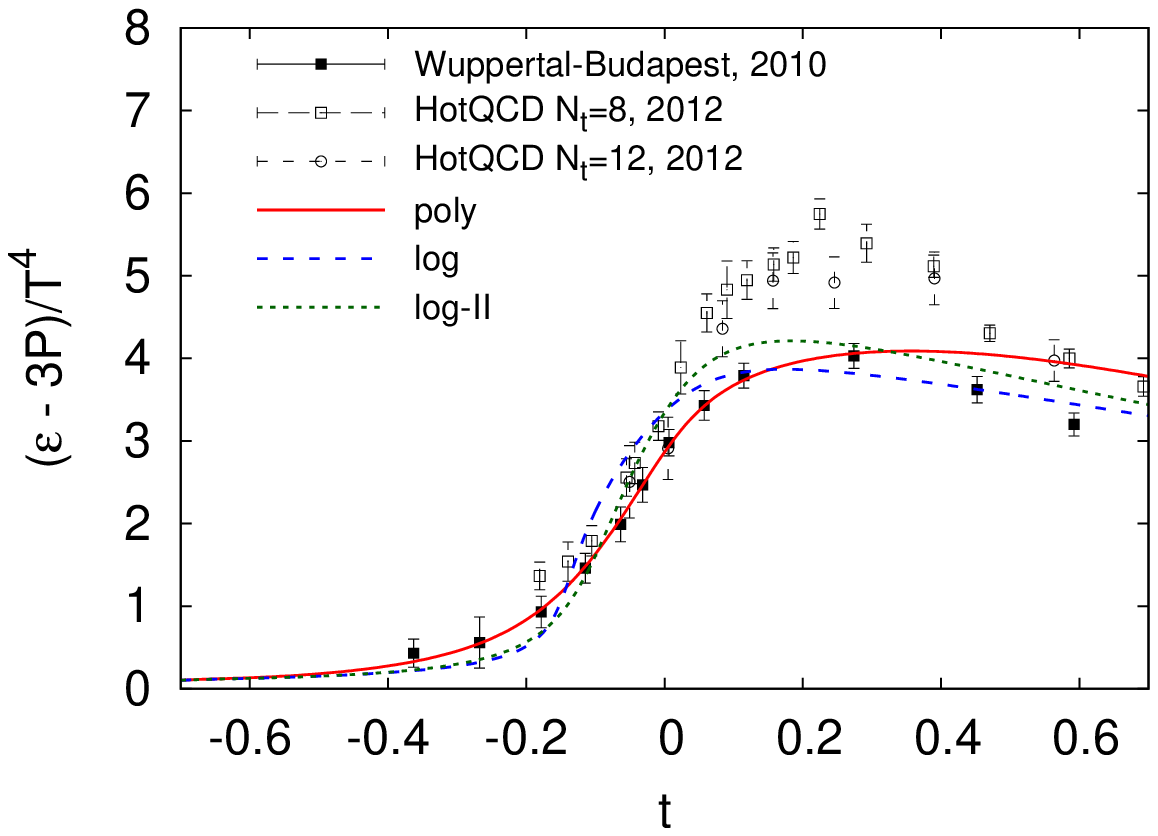}
  \caption{$(2+1)$-flavour FRG results for the pressure (left) and interaction
  measure (right) with polynomial (solid, red)~\cite{Ratti:2005jh}, logarithmic
  (dashed, blue)~\cite{Roessner:2006xn} and enhanced logarithmic
  (dotted, green)~\cite{Lo:2013hla} Polyakov-loop potential with $T_{\rm
  cr}^{\rm  glue}\, =210$ MeV.}
  \label{fig:pdelta_log}
\end{figure*}
%

\section{Conclusion and Outlook}\label{sec:conclusion}

We have investigated order parameters and thermodynamic observables of two and
$2+1$ flavour QCD within effective Polyakov--quark-meson models. This type of
models can be systematically related to full QCD, as e.g.~discussed in
\cite{Herbst:2013ail, Haas:2013qwp}. Thus far, the glue sector of these
models is badly constrained. One often resorts to a Ginzburg--Landau-like ansatz
for the glue potential obtained from fits to lattice Yang-Mills theory.

Recently, first-principles continuum results for the unquenched glue potentials
have become available. These have been used to augment the glue
sector considerably in a mean-field approach to the PQM model
\cite{Haas:2013qwp}. It was shown that by a simple rescaling of the temperature
in the standard Yang-Mills based Polyakov-loop potentials one can already
capture the essential glue dynamics of the unquenched system

In the present work, we have extended the previous investigation 
\cite{Haas:2013qwp} by additionally including thermal and quantum fluctuations
via the functional renormalisation group.
A comparison to lattice QCD simulations with $2+1$ flavours shows excellent
agreement up to temperatures of approximately $1.3$ times the
critical temperature.
Therefore, we conclude that most of the relevant dynamics for the
QCD crossover can already be captured within the PQM model.
Additionally, we have put forward a prediction for the pressure
and interaction measure for two quark flavours, where no recent lattice results
with physical masses are available.

The present work serves as a benchmark of our system at vanishing chemical
potential, which allows us to conclude that we have all relevant fluctuations
included. Since our approach is not restricted to the zero chemical potential
region we can now aim at the full phase diagram, $\mu\geq 0$\,.

\subsection*{Acknowledgements}
We thank L.~Haas, L.~Fister and J.~Schaffner-Bielich for discussions and 
collaboration on related topics. This work is supported by the Helmholtz
Alliance HA216/EMMI, by ERC-AdG-290623, by the FWF through grant P24780-N27 and
Erwin-Schr\"odinger-Stipendium No. J3507, by the
GP-HIR, by the BMBF grant OSPL2VHCTG and by CompStar, a research networking
programme of the European Science Foundation.

\begin{appendix}
\section*{Appendix: Parameter Dependence}\label{app:params}

In this appendix we estimate the parameter dependence of our results.
In particular, we discuss the influence of the Polyakov-loop potential
chosen, our choice of the glue critical temperature $T_{\rm cr}^{\rm glue}$ and
the sigma-meson mass.

\subsection{Polyakov-loop Potential}\label{app:logpot}

\begin{table}
  \begin{tabular}{c|c|c}
         &  $T_\chi$ [MeV] & $T_d$ [MeV]\\\hline\hline
    poly & 172 & 163 \\
    log  & 170 & 146 \\
    log-II & 172 & 156 \\
  \end{tabular}
  \caption{Chiral and deconfinement crossover temperatures for
  different parametrisations of the Polyakov-loop potential, all with $T_{\rm
  cr}^{\rm glue}=210\,\mrm{MeV}$.}
  \label{tab:Tc}
\end{table} 

In the Polyakov-loop extended chiral models one is free to choose a
parametrisation of the Polyakov-loop potential. It is customary to employ a
Landau--Ginzburg-like ansatz and fit the parameters to available lattice data.
However, in this manner only the region close to the minimum is constrained,
not the overall shape. This is the reason why several different functional
forms have been chosen in the past. In practise, when the Polyakov-loop is
coupled to the matter sector, regions away from the minimum are probed and one
should not expect to find exactly the same results with different
versions of the potential.

In the main text we have presented results for a polynomial parametrisation
\cite{Ratti:2005jh}. Here, we want to compare these results to those using a
logarithmic version of the potential~\cite{Roessner:2006xn}
\begin{eqnarray}\label{eq:ulog}
  & & \frac{\mathcal U_{\rm log}(\Phi,\Phibar;t)}{T^4} =
    -\frac{1}{2}a(t)\Phibar\Phi  \\
  & & \qquad \quad +\ b(t) \ln \left[1-6 \Phibar\Phi +
    4\left(\Phi^{3}+\Phibar^{3}\right)
  - 3 \left(\Phibar \Phi\right)^{2}\right]\,,\nonumber
\end{eqnarray}
In this variant, the logarithmic form arises from the integration of
the Haar measure and constrains the Polyakov-loop variables $\Phi,\Phibar$ to
values smaller than one.

Furthermore, we show results with the parametrisation recently proposed in
\cite{Lo:2013hla}
\begin{eqnarray}\label{eq:ulog2}
  & & \frac{\mathcal U_{\rm log-II}(\Phi,\Phibar;t)}{T^4} = \frac{\mathcal 
  U_{\rm log}(\Phi,\Phibar;t)}{T^4}\nonumber \\
  & & \qquad\qquad +\frac{c(t)}{2}  \left(\Phi^3 + \Phibar^3\right) + d(t)
\left(\Phibar\Phi\right)^2\,.
\end{eqnarray}
The parameters of Eqs. \eqref{eq:ulog} and \eqref{eq:ulog2} have been
fixed in \cite{Roessner:2006xn} and \cite{Lo:2013hla}, respectively. 

\tab{tab:Tc} summarises our FRG results for the transition temperatures using
the initial values given in Sec.~\ref{subsec:PQMflow} and these different
Polyakov-loop potentials. Using the polynomial Polyakov-loop potential, the
chiral and deconfinement transitions lie closer to each other than with the
logarithmic ones.

In \fig{fig:pdelta_log}, the pressure (left) and interaction measure
(right) are shown for $N_f=2+1$ and the three parametrisations.
The pressure for the two logarithmic versions lies below the
lattice result at low temperature, while the overall agreement
is quite good also in this case. In the interaction measure, however,
differences are seen more clearly. The trace anomaly resulting from the
standard logarithmic potential rises much steeper than the one
obtained with the polynomial version. The peak height, on the other hand, agrees
well with the results of the Wuppertal-Budapest collaboration, independent of
the parametrisation of the potential.
\begin{figure*}[t]
  \includegraphics[width=.49\textwidth]
  {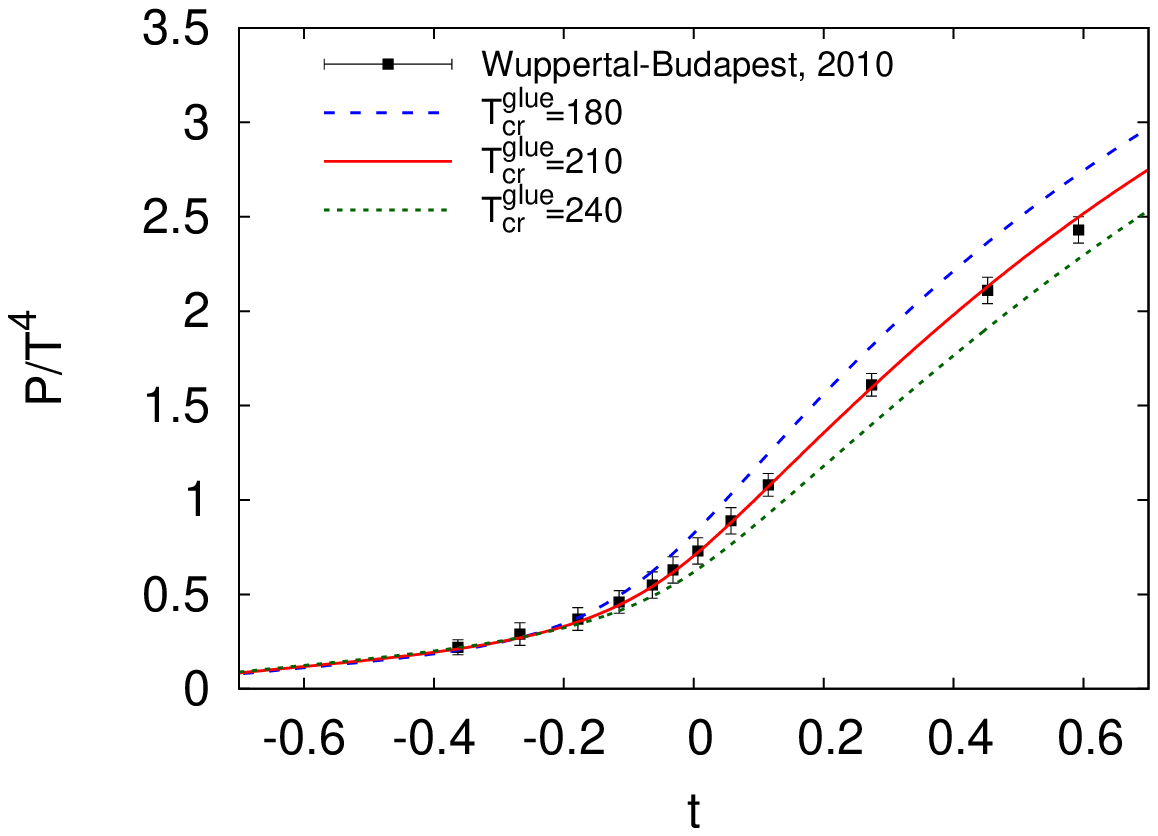}
  \includegraphics[width=.49\textwidth]
  {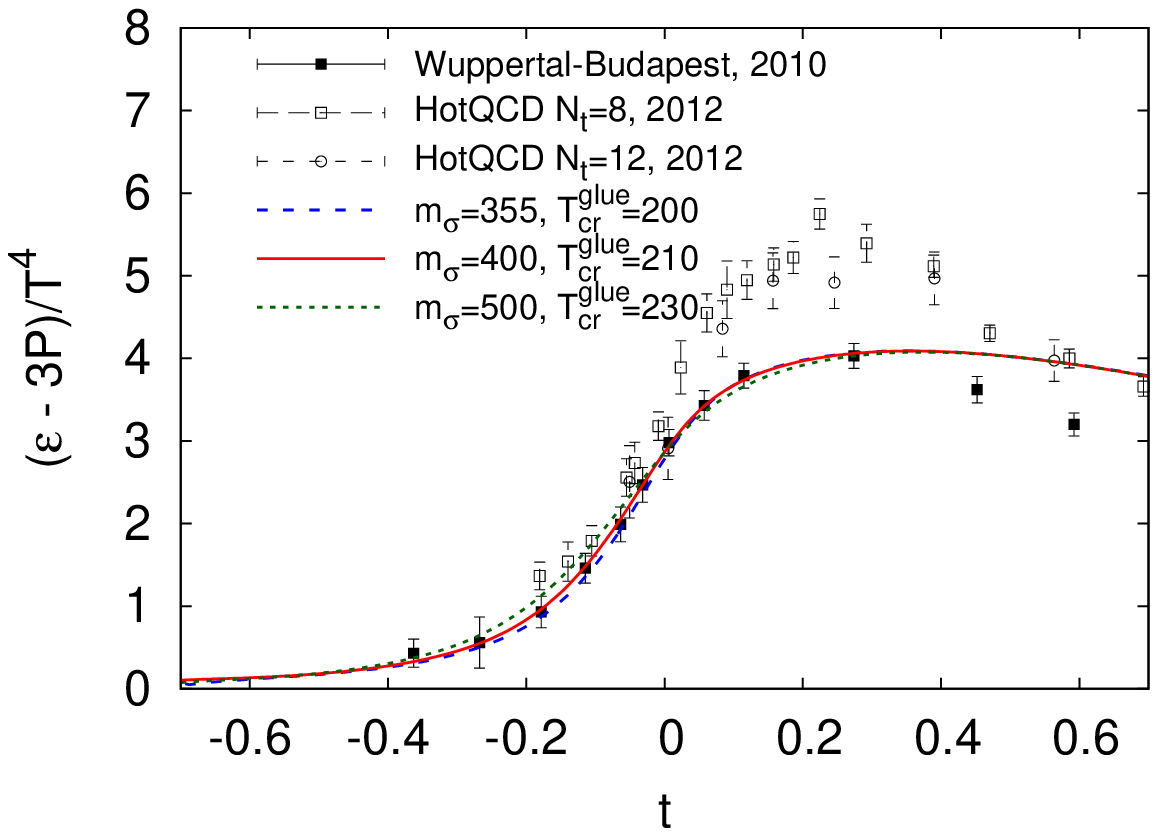}
  \caption{Pressure for different values of $T_{\rm cr}^{\rm glue}$ (left) and
    interaction measure for different choices of $m_\sigma$ with the
    corresponding optimal $T_{\rm cr}^{\rm glue}$ (right).}
  \label{fig:msigma_Tcdep}
\end{figure*}
\begin{table}
  \begin{tabular}{c|c|c}
    $m_\sigma$ [MeV]  & $T_{\chi}$ [MeV]&  $T_{\rm cr}^{\rm glue}$ [MeV]\\
    \hline\hline
    $355$ & $163$ & $200$\\
    $400$ & $172$ & $210$\\
    $500$ & $190$ & $230$
  \end{tabular}
  \caption{Chiral crossover temperatures for different values of the
    sigma-meson mass. In every calculation, the corresponding optimal value
    for $T_{\rm cr}^{\rm glue}$ as obtained from fixing the pressure has been
    used.}
  \label{tab:msigma}
\end{table}
%

\subsection{Glue Critical Temperature}\label{app:Tcr}
Next, we discuss the impact of the glue critical temperature, $T_{\rm
cr}^{\rm glue}$. In the main text we have chosen  $T_{\rm cr}^{\rm
glue}=210$~MeV for $2+1$ flavours with physical anomaly strength.
This value has been obtained by a comparison of the resulting pressure to the
lattice results of the Wuppertal-Budapest collaboration.
\fig{fig:msigma_Tcdep} (left) summarises the dependence of the pressure on
$T_{\rm cr}^{\rm glue}$. In terms of the reduced temperature $t$, the effect of
the glue critical temperature is to stretch the transition region, i.e.~the
transition becomes less steep for larger $T_{\rm cr}^{\rm glue}$.

\subsection{Sigma-Meson Mass}\label{app:msigma}
The sigma meson, $f_0(500)$ is a rather broad resonance, $(400-550) -
i(200-350)$~MeV, which leaves us some freedom to fix this mass in our
setup. The choice of this mass influences, e.g.~the position of the
phase transition and the location of a possible critical endpoint in
the phase diagram \cite{Schaefer:2008hk}.  For $2+1$ flavours we have
fixed $m_\sigma=400$~MeV, see also our discussion in \sec{subsec:crossover}.
In \tab{tab:msigma} we show the critical temperatures for different choices of
the sigma-meson mass. Although the absolute value $T_\chi$ is quite susceptible
to $m_\sigma$, we find that the thermodynamic observables in terms of the
reduced temperature are not. This is demonstrated for the interaction
measure in the right panel of \fig{fig:msigma_Tcdep}. We have to
stress, however, that this $m_\sigma$ independence is partially due to
the fact that a change in the sigma-meson mass can be compensated to
some degree by a change in $T_{\rm cr}^{\rm glue}$, see also
\cite{Haas:2013qwp, Stiele:2013gra}.

\end{appendix}

\bibliography{refs}

\begin{thebibliography}{72}%
\makeatletter
\providecommand \@ifxundefined [1]{%
 \@ifx{#1\undefined}
}%
\providecommand \@ifnum [1]{%
 \ifnum #1\expandafter \@firstoftwo
 \else \expandafter \@secondoftwo
 \fi
}%
\providecommand \@ifx [1]{%
 \ifx #1\expandafter \@firstoftwo
 \else \expandafter \@secondoftwo
 \fi
}%
\providecommand \natexlab [1]{#1}%
\providecommand \enquote  [1]{``#1''}%
\providecommand \bibnamefont  [1]{#1}%
\providecommand \bibfnamefont [1]{#1}%
\providecommand \citenamefont [1]{#1}%
\providecommand \href@noop [0]{\@secondoftwo}%
\providecommand \href [0]{\begingroup \@sanitize@url \@href}%
\providecommand \@href[1]{\@@startlink{#1}\@@href}%
\providecommand \@@href[1]{\endgroup#1\@@endlink}%
\providecommand \@sanitize@url [0]{\catcode `\\12\catcode `\$12\catcode
  `\&12\catcode `\#12\catcode `\^12\catcode `\_12\catcode `\%12\relax}%
\providecommand \@@startlink[1]{}%
\providecommand \@@endlink[0]{}%
\providecommand \url  [0]{\begingroup\@sanitize@url \@url }%
\providecommand \@url [1]{\endgroup\@href {#1}{\urlprefix }}%
\providecommand \urlprefix  [0]{URL }%
\providecommand \Eprint [0]{\href }%
\providecommand \doibase [0]{http://dx.doi.org/}%
\providecommand \selectlanguage [0]{\@gobble}%
\providecommand \bibinfo  [0]{\@secondoftwo}%
\providecommand \bibfield  [0]{\@secondoftwo}%
\providecommand \translation [1]{[#1]}%
\providecommand \BibitemOpen [0]{}%
\providecommand \bibitemStop [0]{}%
\providecommand \bibitemNoStop [0]{.\EOS\space}%
\providecommand \EOS [0]{\spacefactor3000\relax}%
\providecommand \BibitemShut  [1]{\csname bibitem#1\endcsname}%
\let\auto@bib@innerbib\@empty
\bibitem [{\citenamefont {Braun}\ \emph {et~al.}(2011)\citenamefont {Braun},
  \citenamefont {Haas}, \citenamefont {Marhauser},\ and\ \citenamefont
  {Pawlowski}}]{Braun:2009gm}%
  \BibitemOpen
  \bibfield  {author} {\bibinfo {author} {\bibfnamefont {J.}~\bibnamefont
  {Braun}}, \bibinfo {author} {\bibfnamefont {L.~M.}\ \bibnamefont {Haas}},
  \bibinfo {author} {\bibfnamefont {F.}~\bibnamefont {Marhauser}}, \ and\
  \bibinfo {author} {\bibfnamefont {J.~M.}\ \bibnamefont {Pawlowski}},\ }\href
  {\doibase 10.1103/PhysRevLett.106.022002} {\bibfield  {journal} {\bibinfo
  {journal} {Phys.Rev.Lett.}\ }\textbf {\bibinfo {volume} {106}},\ \bibinfo
  {pages} {022002} (\bibinfo {year} {2011})},\ \Eprint
  {http://arxiv.org/abs/0908.0008} {arXiv:0908.0008 [hep-ph]} \BibitemShut
  {NoStop}%
\bibitem [{\citenamefont {Herbst}\ \emph {et~al.}(2013)\citenamefont {Herbst},
  \citenamefont {Pawlowski},\ and\ \citenamefont {Schaefer}}]{Herbst:2013ail}%
  \BibitemOpen
  \bibfield  {author} {\bibinfo {author} {\bibfnamefont {T.~K.}\ \bibnamefont
  {Herbst}}, \bibinfo {author} {\bibfnamefont {J.~M.}\ \bibnamefont
  {Pawlowski}}, \ and\ \bibinfo {author} {\bibfnamefont {B.-J.}\ \bibnamefont
  {Schaefer}},\ }\href {\doibase 10.1103/PhysRevD.88.014007} {\bibfield
  {journal} {\bibinfo  {journal} {Phys. Rev.}\ }\textbf {\bibinfo {volume}
  {D88}},\ \bibinfo {pages} {014007} (\bibinfo {year} {2013})},\ \Eprint
  {http://arxiv.org/abs/1302.1426} {arXiv:1302.1426 [hep-ph]} \BibitemShut
  {NoStop}%
\bibitem [{\citenamefont {Haas}\ \emph {et~al.}(2013)\citenamefont {Haas},
  \citenamefont {Stiele}, \citenamefont {Braun}, \citenamefont {Pawlowski},\
  and\ \citenamefont {Schaffner-Bielich}}]{Haas:2013qwp}%
  \BibitemOpen
  \bibfield  {author} {\bibinfo {author} {\bibfnamefont {L.~M.}\ \bibnamefont
  {Haas}}, \bibinfo {author} {\bibfnamefont {R.}~\bibnamefont {Stiele}},
  \bibinfo {author} {\bibfnamefont {J.}~\bibnamefont {Braun}}, \bibinfo
  {author} {\bibfnamefont {J.~M.}\ \bibnamefont {Pawlowski}}, \ and\ \bibinfo
  {author} {\bibfnamefont {J.}~\bibnamefont {Schaffner-Bielich}},\ }\href
  {\doibase 10.1103/PhysRevD.87.076004} {\bibfield  {journal} {\bibinfo
  {journal} {Phys.Rev.}\ }\textbf {\bibinfo {volume} {D87}},\ \bibinfo {pages}
  {076004} (\bibinfo {year} {2013})},\ \Eprint {http://arxiv.org/abs/1302.1993}
  {arXiv:1302.1993 [hep-ph]} \BibitemShut {NoStop}%
\bibitem [{\citenamefont {Mitter}\ and\ \citenamefont
  {Schaefer}(2013)}]{Mitter:2013fxa}%
  \BibitemOpen
  \bibfield  {author} {\bibinfo {author} {\bibfnamefont {M.}~\bibnamefont
  {Mitter}}\ and\ \bibinfo {author} {\bibfnamefont {B.-J.}\ \bibnamefont
  {Schaefer}},\ }\href@noop {} {\  (\bibinfo {year} {2013})},\ \Eprint
  {http://arxiv.org/abs/1308.3176} {arXiv:1308.3176 [hep-ph]} \BibitemShut
  {NoStop}%
\bibitem [{\citenamefont {Litim}\ and\ \citenamefont
  {Pawlowski}(1999)}]{Litim:1998nf}%
  \BibitemOpen
  \bibfield  {author} {\bibinfo {author} {\bibfnamefont {D.~F.}\ \bibnamefont
  {Litim}}\ and\ \bibinfo {author} {\bibfnamefont {J.~M.}\ \bibnamefont
  {Pawlowski}},\ }\href@noop {} {\bibfield  {journal} {\bibinfo  {journal}
  {World Sci.}\ ,\ \bibinfo {pages} {168}} (\bibinfo {year} {1999})},\ \Eprint
  {http://arxiv.org/abs/hep-th/9901063} {arXiv:hep-th/9901063} \BibitemShut
  {NoStop}%
\bibitem [{\citenamefont {Berges}\ \emph {et~al.}(2002)\citenamefont {Berges},
  \citenamefont {Tetradis},\ and\ \citenamefont {Wetterich}}]{Berges:2000ew}%
  \BibitemOpen
  \bibfield  {author} {\bibinfo {author} {\bibfnamefont {J.}~\bibnamefont
  {Berges}}, \bibinfo {author} {\bibfnamefont {N.}~\bibnamefont {Tetradis}}, \
  and\ \bibinfo {author} {\bibfnamefont {C.}~\bibnamefont {Wetterich}},\ }\href
  {\doibase 10.1016/S0370-1573(01)00098-9} {\bibfield  {journal} {\bibinfo
  {journal} {Phys.Rept.}\ }\textbf {\bibinfo {volume} {363}},\ \bibinfo {pages}
  {223} (\bibinfo {year} {2002})},\ \Eprint
  {http://arxiv.org/abs/hep-ph/0005122} {arXiv:hep-ph/0005122} \BibitemShut
  {NoStop}%
\bibitem [{\citenamefont {Pawlowski}(2007)}]{Pawlowski:2005xe}%
  \BibitemOpen
  \bibfield  {author} {\bibinfo {author} {\bibfnamefont {J.~M.}\ \bibnamefont
  {Pawlowski}},\ }\href {\doibase 10.1016/j.aop.2007.01.007} {\bibfield
  {journal} {\bibinfo  {journal} {Annals Phys.}\ }\textbf {\bibinfo {volume}
  {322}},\ \bibinfo {pages} {2831} (\bibinfo {year} {2007})},\ \Eprint
  {http://arxiv.org/abs/hep-th/0512261} {arXiv:hep-th/0512261} \BibitemShut
  {NoStop}%
\bibitem [{\citenamefont {Gies}(2012)}]{Gies:2006wv}%
  \BibitemOpen
  \bibfield  {author} {\bibinfo {author} {\bibfnamefont {H.}~\bibnamefont
  {Gies}},\ }\href {\doibase 10.1007/978-3-642-27320-9_6} {\bibfield  {journal}
  {\bibinfo  {journal} {Lect.Notes Phys.}\ }\textbf {\bibinfo {volume} {852}},\
  \bibinfo {pages} {287} (\bibinfo {year} {2012})},\ \Eprint
  {http://arxiv.org/abs/hep-ph/0611146} {arXiv:hep-ph/0611146} \BibitemShut
  {NoStop}%
\bibitem [{\citenamefont {Schaefer}\ and\ \citenamefont
  {Wambach}(2008)}]{Schaefer:2006sr}%
  \BibitemOpen
  \bibfield  {author} {\bibinfo {author} {\bibfnamefont {B.-J.}\ \bibnamefont
  {Schaefer}}\ and\ \bibinfo {author} {\bibfnamefont {J.}~\bibnamefont
  {Wambach}},\ }\href {\doibase 10.1134/S1063779608070083} {\bibfield
  {journal} {\bibinfo  {journal} {Phys.Part.Nucl.}\ }\textbf {\bibinfo {volume}
  {39}},\ \bibinfo {pages} {1025} (\bibinfo {year} {2008})},\ \Eprint
  {http://arxiv.org/abs/hep-ph/0611191} {arXiv:hep-ph/0611191} \BibitemShut
  {NoStop}%
\bibitem [{\citenamefont {Pawlowski}(2011)}]{Pawlowski:2010ht}%
  \BibitemOpen
  \bibfield  {author} {\bibinfo {author} {\bibfnamefont {J.~M.}\ \bibnamefont
  {Pawlowski}},\ }\href {\doibase 10.1063/1.3574945} {\bibfield  {journal}
  {\bibinfo  {journal} {AIP Conf.Proc.}\ }\textbf {\bibinfo {volume} {1343}},\
  \bibinfo {pages} {75} (\bibinfo {year} {2011})},\ \Eprint
  {http://arxiv.org/abs/1012.5075} {arXiv:1012.5075 [hep-ph]} \BibitemShut
  {NoStop}%
\bibitem [{\citenamefont {Braun}(2012)}]{Braun:2011pp}%
  \BibitemOpen
  \bibfield  {author} {\bibinfo {author} {\bibfnamefont {J.}~\bibnamefont
  {Braun}},\ }\href {\doibase 10.1088/0954-3899/39/3/033001} {\bibfield
  {journal} {\bibinfo  {journal} {J.Phys.}\ }\textbf {\bibinfo {volume}
  {G39}},\ \bibinfo {pages} {033001} (\bibinfo {year} {2012})},\ \Eprint
  {http://arxiv.org/abs/1108.4449} {arXiv:1108.4449 [hep-ph]} \BibitemShut
  {NoStop}%
\bibitem [{\citenamefont {von Smekal}(2012)}]{vonSmekal:2012vx}%
  \BibitemOpen
  \bibfield  {author} {\bibinfo {author} {\bibfnamefont {L.}~\bibnamefont {von
  Smekal}},\ }\href {\doibase 10.1016/j.nuclphysbps.2012.06.006} {\bibfield
  {journal} {\bibinfo  {journal} {Nucl.Phys.Proc.Suppl.}\ }\textbf {\bibinfo
  {volume} {228}},\ \bibinfo {pages} {179} (\bibinfo {year} {2012})},\ \Eprint
  {http://arxiv.org/abs/1205.4205} {arXiv:1205.4205 [hep-ph]} \BibitemShut
  {NoStop}%
\bibitem [{\citenamefont {Braun}\ \emph
  {et~al.}(2010{\natexlab{a}})\citenamefont {Braun}, \citenamefont {Gies},\
  and\ \citenamefont {Pawlowski}}]{Braun:2007bx}%
  \BibitemOpen
  \bibfield  {author} {\bibinfo {author} {\bibfnamefont {J.}~\bibnamefont
  {Braun}}, \bibinfo {author} {\bibfnamefont {H.}~\bibnamefont {Gies}}, \ and\
  \bibinfo {author} {\bibfnamefont {J.~M.}\ \bibnamefont {Pawlowski}},\ }\href
  {\doibase 10.1016/j.physletb.2010.01.009} {\bibfield  {journal} {\bibinfo
  {journal} {Phys.Lett.}\ }\textbf {\bibinfo {volume} {B684}},\ \bibinfo
  {pages} {262} (\bibinfo {year} {2010}{\natexlab{a}})},\ \Eprint
  {http://arxiv.org/abs/0708.2413} {arXiv:0708.2413 [hep-th]} \BibitemShut
  {NoStop}%
\bibitem [{\citenamefont {Fischer}\ and\ \citenamefont
  {Luecker}(2013)}]{Fischer:2012vc}%
  \BibitemOpen
  \bibfield  {author} {\bibinfo {author} {\bibfnamefont {C.~S.}\ \bibnamefont
  {Fischer}}\ and\ \bibinfo {author} {\bibfnamefont {J.}~\bibnamefont
  {Luecker}},\ }\href {\doibase 10.1016/j.physletb.2012.11.054} {\bibfield
  {journal} {\bibinfo  {journal} {Phys.Lett.}\ }\textbf {\bibinfo {volume}
  {B718}},\ \bibinfo {pages} {1036} (\bibinfo {year} {2013})},\ \Eprint
  {http://arxiv.org/abs/1206.5191} {arXiv:1206.5191 [hep-ph]} \BibitemShut
  {NoStop}%
\bibitem [{\citenamefont {Fister}\ and\ \citenamefont
  {Pawlowski}(2013)}]{Fister:2013bh}%
  \BibitemOpen
  \bibfield  {author} {\bibinfo {author} {\bibfnamefont {L.}~\bibnamefont
  {Fister}}\ and\ \bibinfo {author} {\bibfnamefont {J.~M.}\ \bibnamefont
  {Pawlowski}},\ }\href {\doibase 10.1103/PhysRevD.88.045010} {\bibfield
  {journal} {\bibinfo  {journal} {Phys.Rev.}\ }\textbf {\bibinfo {volume}
  {D88}},\ \bibinfo {pages} {045010} (\bibinfo {year} {2013})},\ \Eprint
  {http://arxiv.org/abs/1301.4163} {arXiv:1301.4163 [hep-ph]} \BibitemShut
  {NoStop}%
\bibitem [{\citenamefont {Fischer}\ \emph {et~al.}(2013)\citenamefont
  {Fischer}, \citenamefont {Fister}, \citenamefont {Luecker},\ and\
  \citenamefont {Pawlowski}}]{Fischer:2013eca}%
  \BibitemOpen
  \bibfield  {author} {\bibinfo {author} {\bibfnamefont {C.~S.}\ \bibnamefont
  {Fischer}}, \bibinfo {author} {\bibfnamefont {L.}~\bibnamefont {Fister}},
  \bibinfo {author} {\bibfnamefont {J.}~\bibnamefont {Luecker}}, \ and\
  \bibinfo {author} {\bibfnamefont {J.~M.}\ \bibnamefont {Pawlowski}},\
  }\href@noop {} {\  (\bibinfo {year} {2013})},\ \Eprint
  {http://arxiv.org/abs/1306.6022} {arXiv:1306.6022 [hep-ph]} \BibitemShut
  {NoStop}%
\bibitem [{\citenamefont {Meisinger}\ and\ \citenamefont
  {Ogilvie}(1996)}]{Meisinger:1995ih}%
  \BibitemOpen
  \bibfield  {author} {\bibinfo {author} {\bibfnamefont {P.~N.}\ \bibnamefont
  {Meisinger}}\ and\ \bibinfo {author} {\bibfnamefont {M.~C.}\ \bibnamefont
  {Ogilvie}},\ }\href {\doibase 10.1016/0370-2693(96)00447-9} {\bibfield
  {journal} {\bibinfo  {journal} {Phys.Lett.}\ }\textbf {\bibinfo {volume}
  {B379}},\ \bibinfo {pages} {163} (\bibinfo {year} {1996})},\ \Eprint
  {http://arxiv.org/abs/hep-lat/9512011} {arXiv:hep-lat/9512011} \BibitemShut
  {NoStop}%
\bibitem [{\citenamefont {Pisarski}(2000)}]{Pisarski:2000eq}%
  \BibitemOpen
  \bibfield  {author} {\bibinfo {author} {\bibfnamefont {R.~D.}\ \bibnamefont
  {Pisarski}},\ }\href {\doibase 10.1103/PhysRevD.62.111501} {\bibfield
  {journal} {\bibinfo  {journal} {Phys.Rev.}\ }\textbf {\bibinfo {volume}
  {D62}},\ \bibinfo {pages} {111501} (\bibinfo {year} {2000})},\ \Eprint
  {http://arxiv.org/abs/hep-ph/0006205} {arXiv:hep-ph/0006205} \BibitemShut
  {NoStop}%
\bibitem [{\citenamefont {Fukushima}(2004)}]{Fukushima:2003fw}%
  \BibitemOpen
  \bibfield  {author} {\bibinfo {author} {\bibfnamefont {K.}~\bibnamefont
  {Fukushima}},\ }\href {\doibase 10.1016/j.physletb.2004.04.027} {\bibfield
  {journal} {\bibinfo  {journal} {Phys.Lett.}\ }\textbf {\bibinfo {volume}
  {B591}},\ \bibinfo {pages} {277} (\bibinfo {year} {2004})},\ \Eprint
  {http://arxiv.org/abs/hep-ph/0310121} {arXiv:hep-ph/0310121} \BibitemShut
  {NoStop}%
\bibitem [{\citenamefont {Megias}\ \emph {et~al.}(2006)\citenamefont {Megias},
  \citenamefont {Ruiz~Arriola},\ and\ \citenamefont {Salcedo}}]{Megias:2004hj}%
  \BibitemOpen
  \bibfield  {author} {\bibinfo {author} {\bibfnamefont {E.}~\bibnamefont
  {Megias}}, \bibinfo {author} {\bibfnamefont {E.}~\bibnamefont
  {Ruiz~Arriola}}, \ and\ \bibinfo {author} {\bibfnamefont {L.}~\bibnamefont
  {Salcedo}},\ }\href {\doibase 10.1103/PhysRevD.74.065005} {\bibfield
  {journal} {\bibinfo  {journal} {Phys.Rev.}\ }\textbf {\bibinfo {volume}
  {D74}},\ \bibinfo {pages} {065005} (\bibinfo {year} {2006})},\ \Eprint
  {http://arxiv.org/abs/hep-ph/0412308} {arXiv:hep-ph/0412308} \BibitemShut
  {NoStop}%
\bibitem [{\citenamefont {Ratti}\ \emph {et~al.}(2006)\citenamefont {Ratti},
  \citenamefont {Thaler},\ and\ \citenamefont {Weise}}]{Ratti:2005jh}%
  \BibitemOpen
  \bibfield  {author} {\bibinfo {author} {\bibfnamefont {C.}~\bibnamefont
  {Ratti}}, \bibinfo {author} {\bibfnamefont {M.~A.}\ \bibnamefont {Thaler}}, \
  and\ \bibinfo {author} {\bibfnamefont {W.}~\bibnamefont {Weise}},\ }\href
  {\doibase 10.1103/PhysRevD.73.014019} {\bibfield  {journal} {\bibinfo
  {journal} {Phys.Rev.}\ }\textbf {\bibinfo {volume} {D73}},\ \bibinfo {pages}
  {014019} (\bibinfo {year} {2006})},\ \Eprint
  {http://arxiv.org/abs/hep-ph/0506234} {arXiv:hep-ph/0506234} \BibitemShut
  {NoStop}%
\bibitem [{\citenamefont {Mukherjee}\ \emph {et~al.}(2007)\citenamefont
  {Mukherjee}, \citenamefont {Mustafa},\ and\ \citenamefont
  {Ray}}]{Mukherjee:2006hq}%
  \BibitemOpen
  \bibfield  {author} {\bibinfo {author} {\bibfnamefont {S.}~\bibnamefont
  {Mukherjee}}, \bibinfo {author} {\bibfnamefont {M.~G.}\ \bibnamefont
  {Mustafa}}, \ and\ \bibinfo {author} {\bibfnamefont {R.}~\bibnamefont
  {Ray}},\ }\href {\doibase 10.1103/PhysRevD.75.094015} {\bibfield  {journal}
  {\bibinfo  {journal} {Phys.Rev.}\ }\textbf {\bibinfo {volume} {D75}},\
  \bibinfo {pages} {094015} (\bibinfo {year} {2007})},\ \Eprint
  {http://arxiv.org/abs/hep-ph/0609249} {arXiv:hep-ph/0609249} \BibitemShut
  {NoStop}%
\bibitem [{\citenamefont {Roessner}\ \emph {et~al.}(2007)\citenamefont
  {Roessner}, \citenamefont {Ratti},\ and\ \citenamefont
  {Weise}}]{Roessner:2006xn}%
  \BibitemOpen
  \bibfield  {author} {\bibinfo {author} {\bibfnamefont {S.}~\bibnamefont
  {Roessner}}, \bibinfo {author} {\bibfnamefont {C.}~\bibnamefont {Ratti}}, \
  and\ \bibinfo {author} {\bibfnamefont {W.}~\bibnamefont {Weise}},\ }\href
  {\doibase 10.1103/PhysRevD.75.034007} {\bibfield  {journal} {\bibinfo
  {journal} {Phys.Rev.}\ }\textbf {\bibinfo {volume} {D75}},\ \bibinfo {pages}
  {034007} (\bibinfo {year} {2007})},\ \Eprint
  {http://arxiv.org/abs/hep-ph/0609281} {arXiv:hep-ph/0609281} \BibitemShut
  {NoStop}%
\bibitem [{\citenamefont {Sasaki}\ \emph {et~al.}(2007)\citenamefont {Sasaki},
  \citenamefont {Friman},\ and\ \citenamefont {Redlich}}]{Sasaki:2006ww}%
  \BibitemOpen
  \bibfield  {author} {\bibinfo {author} {\bibfnamefont {C.}~\bibnamefont
  {Sasaki}}, \bibinfo {author} {\bibfnamefont {B.}~\bibnamefont {Friman}}, \
  and\ \bibinfo {author} {\bibfnamefont {K.}~\bibnamefont {Redlich}},\ }\href
  {\doibase 10.1103/PhysRevD.75.074013} {\bibfield  {journal} {\bibinfo
  {journal} {Phys. Rev.}\ }\textbf {\bibinfo {volume} {D75}},\ \bibinfo {pages}
  {074013} (\bibinfo {year} {2007})},\ \Eprint
  {http://arxiv.org/abs/hep-ph/0611147} {arXiv:hep-ph/0611147} \BibitemShut
  {NoStop}%
\bibitem [{\citenamefont {Schaefer}\ \emph {et~al.}(2007)\citenamefont
  {Schaefer}, \citenamefont {Pawlowski},\ and\ \citenamefont
  {Wambach}}]{Schaefer:2007pw}%
  \BibitemOpen
  \bibfield  {author} {\bibinfo {author} {\bibfnamefont {B.-J.}\ \bibnamefont
  {Schaefer}}, \bibinfo {author} {\bibfnamefont {J.~M.}\ \bibnamefont
  {Pawlowski}}, \ and\ \bibinfo {author} {\bibfnamefont {J.}~\bibnamefont
  {Wambach}},\ }\href {\doibase 10.1103/PhysRevD.76.074023} {\bibfield
  {journal} {\bibinfo  {journal} {Phys.Rev.}\ }\textbf {\bibinfo {volume}
  {D76}},\ \bibinfo {pages} {074023} (\bibinfo {year} {2007})},\ \Eprint
  {http://arxiv.org/abs/0704.3234} {arXiv:0704.3234 [hep-ph]} \BibitemShut
  {NoStop}%
\bibitem [{\citenamefont {Fraga}\ and\ \citenamefont
  {Mocsy}(2007)}]{Fraga:2007un}%
  \BibitemOpen
  \bibfield  {author} {\bibinfo {author} {\bibfnamefont {E.}~\bibnamefont
  {Fraga}}\ and\ \bibinfo {author} {\bibfnamefont {A.}~\bibnamefont {Mocsy}},\
  }\href {\doibase 10.1590/S0103-97332007000200020} {\bibfield  {journal}
  {\bibinfo  {journal} {Braz.J.Phys.}\ }\textbf {\bibinfo {volume} {37}},\
  \bibinfo {pages} {281} (\bibinfo {year} {2007})},\ \Eprint
  {http://arxiv.org/abs/hep-ph/0701102} {arXiv:hep-ph/0701102} \BibitemShut
  {NoStop}%
\bibitem [{\citenamefont {Fukushima}(2008)}]{Fukushima:2008wg}%
  \BibitemOpen
  \bibfield  {author} {\bibinfo {author} {\bibfnamefont {K.}~\bibnamefont
  {Fukushima}},\ }\href {\doibase 10.1103/PhysRevD.77.114028,
  10.1103/PhysRevD.78.039902} {\bibfield  {journal} {\bibinfo  {journal}
  {Phys.Rev.}\ }\textbf {\bibinfo {volume} {D77}},\ \bibinfo {pages} {114028}
  (\bibinfo {year} {2008})},\ \Eprint {http://arxiv.org/abs/0803.3318}
  {arXiv:0803.3318 [hep-ph]} \BibitemShut {NoStop}%
\bibitem [{\citenamefont {Sakai}\ \emph {et~al.}(2008)\citenamefont {Sakai},
  \citenamefont {Kashiwa}, \citenamefont {Kouno},\ and\ \citenamefont
  {Yahiro}}]{Sakai:2008py}%
  \BibitemOpen
  \bibfield  {author} {\bibinfo {author} {\bibfnamefont {Y.}~\bibnamefont
  {Sakai}}, \bibinfo {author} {\bibfnamefont {K.}~\bibnamefont {Kashiwa}},
  \bibinfo {author} {\bibfnamefont {H.}~\bibnamefont {Kouno}}, \ and\ \bibinfo
  {author} {\bibfnamefont {M.}~\bibnamefont {Yahiro}},\ }\href {\doibase
  10.1103/PhysRevD.77.051901} {\bibfield  {journal} {\bibinfo  {journal}
  {Phys.Rev.}\ }\textbf {\bibinfo {volume} {D77}},\ \bibinfo {pages} {051901}
  (\bibinfo {year} {2008})},\ \Eprint {http://arxiv.org/abs/0801.0034}
  {arXiv:0801.0034 [hep-ph]} \BibitemShut {NoStop}%
\bibitem [{\citenamefont {Herbst}\ \emph {et~al.}(2011)\citenamefont {Herbst},
  \citenamefont {Pawlowski},\ and\ \citenamefont {Schaefer}}]{Herbst:2010rf}%
  \BibitemOpen
  \bibfield  {author} {\bibinfo {author} {\bibfnamefont {T.~K.}\ \bibnamefont
  {Herbst}}, \bibinfo {author} {\bibfnamefont {J.~M.}\ \bibnamefont
  {Pawlowski}}, \ and\ \bibinfo {author} {\bibfnamefont {B.-J.}\ \bibnamefont
  {Schaefer}},\ }\href {\doibase 10.1016/j.physletb.2010.12.003} {\bibfield
  {journal} {\bibinfo  {journal} {Phys.Lett.}\ }\textbf {\bibinfo {volume}
  {B696}},\ \bibinfo {pages} {58} (\bibinfo {year} {2011})},\ \Eprint
  {http://arxiv.org/abs/1008.0081} {arXiv:1008.0081 [hep-ph]} \BibitemShut
  {NoStop}%
\bibitem [{\citenamefont {Skokov}\ \emph {et~al.}(2011)\citenamefont {Skokov},
  \citenamefont {Friman},\ and\ \citenamefont {Redlich}}]{Skokov:2010uh}%
  \BibitemOpen
  \bibfield  {author} {\bibinfo {author} {\bibfnamefont {V.}~\bibnamefont
  {Skokov}}, \bibinfo {author} {\bibfnamefont {B.}~\bibnamefont {Friman}}, \
  and\ \bibinfo {author} {\bibfnamefont {K.}~\bibnamefont {Redlich}},\ }\href
  {\doibase 10.1103/PhysRevC.83.054904} {\bibfield  {journal} {\bibinfo
  {journal} {Phys.Rev.}\ }\textbf {\bibinfo {volume} {C83}},\ \bibinfo {pages}
  {054904} (\bibinfo {year} {2011})},\ \Eprint {http://arxiv.org/abs/1008.4570}
  {arXiv:1008.4570 [hep-ph]} \BibitemShut {NoStop}%
\bibitem [{\citenamefont {Skokov}\ \emph
  {et~al.}(2010{\natexlab{a}})\citenamefont {Skokov}, \citenamefont {Stokic},
  \citenamefont {Friman},\ and\ \citenamefont {Redlich}}]{Skokov:2010wb}%
  \BibitemOpen
  \bibfield  {author} {\bibinfo {author} {\bibfnamefont {V.}~\bibnamefont
  {Skokov}}, \bibinfo {author} {\bibfnamefont {B.}~\bibnamefont {Stokic}},
  \bibinfo {author} {\bibfnamefont {B.}~\bibnamefont {Friman}}, \ and\ \bibinfo
  {author} {\bibfnamefont {K.}~\bibnamefont {Redlich}},\ }\href {\doibase
  10.1103/PhysRevC.82.015206} {\bibfield  {journal} {\bibinfo  {journal} {Phys.
  Rev.}\ }\textbf {\bibinfo {volume} {C82}},\ \bibinfo {pages} {015206}
  (\bibinfo {year} {2010}{\natexlab{a}})},\ \Eprint
  {http://arxiv.org/abs/1004.2665} {arXiv:1004.2665 [hep-ph]} \BibitemShut
  {NoStop}%
\bibitem [{\citenamefont {Schaefer}\ and\ \citenamefont
  {Wagner}(2012)}]{Schaefer:2011ex}%
  \BibitemOpen
  \bibfield  {author} {\bibinfo {author} {\bibfnamefont {B.-J.}\ \bibnamefont
  {Schaefer}}\ and\ \bibinfo {author} {\bibfnamefont {M.}~\bibnamefont
  {Wagner}},\ }\href {\doibase 10.1103/PhysRevD.85.034027} {\bibfield
  {journal} {\bibinfo  {journal} {Phys.Rev.}\ }\textbf {\bibinfo {volume}
  {D85}},\ \bibinfo {pages} {034027} (\bibinfo {year} {2012})},\ \Eprint
  {http://arxiv.org/abs/1111.6871} {arXiv:1111.6871 [hep-ph]} \BibitemShut
  {NoStop}%
\bibitem [{\citenamefont {Braun}\ and\ \citenamefont
  {Janot}(2011)}]{Braun:2011fw}%
  \BibitemOpen
  \bibfield  {author} {\bibinfo {author} {\bibfnamefont {J.}~\bibnamefont
  {Braun}}\ and\ \bibinfo {author} {\bibfnamefont {A.}~\bibnamefont {Janot}},\
  }\href {\doibase 10.1103/PhysRevD.84.114022} {\bibfield  {journal} {\bibinfo
  {journal} {Phys.Rev.}\ }\textbf {\bibinfo {volume} {D84}},\ \bibinfo {pages}
  {114022} (\bibinfo {year} {2011})},\ \Eprint {http://arxiv.org/abs/1102.4841}
  {arXiv:1102.4841 [hep-ph]} \BibitemShut {NoStop}%
\bibitem [{\citenamefont {Kamikado}\ \emph {et~al.}(2013)\citenamefont
  {Kamikado}, \citenamefont {Strodthoff}, \citenamefont {von Smekal},\ and\
  \citenamefont {Wambach}}]{Kamikado:2012bt}%
  \BibitemOpen
  \bibfield  {author} {\bibinfo {author} {\bibfnamefont {K.}~\bibnamefont
  {Kamikado}}, \bibinfo {author} {\bibfnamefont {N.}~\bibnamefont
  {Strodthoff}}, \bibinfo {author} {\bibfnamefont {L.}~\bibnamefont {von
  Smekal}}, \ and\ \bibinfo {author} {\bibfnamefont {J.}~\bibnamefont
  {Wambach}},\ }\href {\doibase 10.1016/j.physletb.2012.11.055} {\bibfield
  {journal} {\bibinfo  {journal} {Phys.Lett.}\ }\textbf {\bibinfo {volume}
  {B718}},\ \bibinfo {pages} {1044} (\bibinfo {year} {2013})},\ \Eprint
  {http://arxiv.org/abs/1207.0400} {arXiv:1207.0400 [hep-ph]} \BibitemShut
  {NoStop}%
\bibitem [{\citenamefont {Braun}\ and\ \citenamefont
  {Herbst}(2012)}]{Braun:2012zq}%
  \BibitemOpen
  \bibfield  {author} {\bibinfo {author} {\bibfnamefont {J.}~\bibnamefont
  {Braun}}\ and\ \bibinfo {author} {\bibfnamefont {T.~K.}\ \bibnamefont
  {Herbst}},\ }\href@noop {} {\  (\bibinfo {year} {2012})},\ \Eprint
  {http://arxiv.org/abs/1205.0779} {arXiv:1205.0779 [hep-ph]} \BibitemShut
  {NoStop}%
\bibitem [{\citenamefont {Fukushima}\ and\ \citenamefont
  {Kashiwa}(2013)}]{Fukushima:2012qa}%
  \BibitemOpen
  \bibfield  {author} {\bibinfo {author} {\bibfnamefont {K.}~\bibnamefont
  {Fukushima}}\ and\ \bibinfo {author} {\bibfnamefont {K.}~\bibnamefont
  {Kashiwa}},\ }\href {\doibase 10.1016/j.physletb.2013.05.037} {\bibfield
  {journal} {\bibinfo  {journal} {Phys.Lett.}\ }\textbf {\bibinfo {volume}
  {B723}},\ \bibinfo {pages} {360} (\bibinfo {year} {2013})},\ \Eprint
  {http://arxiv.org/abs/1206.0685} {arXiv:1206.0685 [hep-ph]} \BibitemShut
  {NoStop}%
\bibitem [{\citenamefont {Mintz}\ \emph {et~al.}(2013)\citenamefont {Mintz},
  \citenamefont {Stiele}, \citenamefont {Ramos},\ and\ \citenamefont
  {Schaffner-Bielich}}]{Mintz:2012mz}%
  \BibitemOpen
  \bibfield  {author} {\bibinfo {author} {\bibfnamefont {B.~W.}\ \bibnamefont
  {Mintz}}, \bibinfo {author} {\bibfnamefont {R.}~\bibnamefont {Stiele}},
  \bibinfo {author} {\bibfnamefont {R.~O.}\ \bibnamefont {Ramos}}, \ and\
  \bibinfo {author} {\bibfnamefont {J.}~\bibnamefont {Schaffner-Bielich}},\
  }\href {\doibase 10.1103/PhysRevD.87.036004} {\bibfield  {journal} {\bibinfo
  {journal} {Phys.Rev.}\ }\textbf {\bibinfo {volume} {D87}},\ \bibinfo {pages}
  {036004} (\bibinfo {year} {2013})},\ \Eprint {http://arxiv.org/abs/1212.1184}
  {arXiv:1212.1184 [hep-ph]} \BibitemShut {NoStop}%
\bibitem [{\citenamefont {Stiele}\ \emph {et~al.}(2013)\citenamefont {Stiele},
  \citenamefont {Fraga},\ and\ \citenamefont
  {Schaffner-Bielich}}]{Stiele:2013pma}%
  \BibitemOpen
  \bibfield  {author} {\bibinfo {author} {\bibfnamefont {R.}~\bibnamefont
  {Stiele}}, \bibinfo {author} {\bibfnamefont {E.~S.}\ \bibnamefont {Fraga}}, \
  and\ \bibinfo {author} {\bibfnamefont {J.}~\bibnamefont
  {Schaffner-Bielich}},\ }\href@noop {} {\  (\bibinfo {year} {2013})},\ \Eprint
  {http://arxiv.org/abs/1307.2851} {arXiv:1307.2851 [hep-ph]} \BibitemShut
  {NoStop}%
\bibitem [{\citenamefont {Strodthoff}\ and\ \citenamefont {von
  Smekal}(2013)}]{Strodthoff:2013cua}%
  \BibitemOpen
  \bibfield  {author} {\bibinfo {author} {\bibfnamefont {N.}~\bibnamefont
  {Strodthoff}}\ and\ \bibinfo {author} {\bibfnamefont {L.}~\bibnamefont {von
  Smekal}},\ }\href@noop {} {\  (\bibinfo {year} {2013})},\ \Eprint
  {http://arxiv.org/abs/1306.2897} {arXiv:1306.2897 [hep-ph]} \BibitemShut
  {NoStop}%
\bibitem [{\citenamefont {Stiele}\ \emph {et~al.}(2012)\citenamefont {Stiele},
  \citenamefont {Haas}, \citenamefont {Braun}, \citenamefont {Pawlowski},\ and\
  \citenamefont {Schaffner-Bielich}}]{Stiele:2013gra}%
  \BibitemOpen
  \bibfield  {author} {\bibinfo {author} {\bibfnamefont {R.}~\bibnamefont
  {Stiele}}, \bibinfo {author} {\bibfnamefont {L.~M.}\ \bibnamefont {Haas}},
  \bibinfo {author} {\bibfnamefont {J.}~\bibnamefont {Braun}}, \bibinfo
  {author} {\bibfnamefont {J.~M.}\ \bibnamefont {Pawlowski}}, \ and\ \bibinfo
  {author} {\bibfnamefont {J.}~\bibnamefont {Schaffner-Bielich}},\ }\href@noop
  {} {\bibfield  {journal} {\bibinfo  {journal} {PoS}\ }\textbf {\bibinfo
  {volume} {ConfinementX}},\ \bibinfo {pages} {215} (\bibinfo {year} {2012})},\
  \Eprint {http://arxiv.org/abs/1303.3742} {arXiv:1303.3742 [hep-ph]}
  \BibitemShut {NoStop}%
\bibitem [{\citenamefont {Gies}\ and\ \citenamefont
  {Wetterich}(2002)}]{Gies:2001nw}%
  \BibitemOpen
  \bibfield  {author} {\bibinfo {author} {\bibfnamefont {H.}~\bibnamefont
  {Gies}}\ and\ \bibinfo {author} {\bibfnamefont {C.}~\bibnamefont
  {Wetterich}},\ }\href {\doibase 10.1103/PhysRevD.65.065001} {\bibfield
  {journal} {\bibinfo  {journal} {Phys.Rev.}\ }\textbf {\bibinfo {volume}
  {D65}},\ \bibinfo {pages} {065001} (\bibinfo {year} {2002})},\ \Eprint
  {http://arxiv.org/abs/hep-th/0107221} {arXiv:hep-th/0107221} \BibitemShut
  {NoStop}%
\bibitem [{\citenamefont {Gies}\ and\ \citenamefont
  {Wetterich}(2004)}]{Gies:2002hq}%
  \BibitemOpen
  \bibfield  {author} {\bibinfo {author} {\bibfnamefont {H.}~\bibnamefont
  {Gies}}\ and\ \bibinfo {author} {\bibfnamefont {C.}~\bibnamefont
  {Wetterich}},\ }\href {\doibase 10.1103/PhysRevD.69.025001} {\bibfield
  {journal} {\bibinfo  {journal} {Phys.Rev.}\ }\textbf {\bibinfo {volume}
  {D69}},\ \bibinfo {pages} {025001} (\bibinfo {year} {2004})},\ \Eprint
  {http://arxiv.org/abs/hep-th/0209183} {arXiv:hep-th/0209183} \BibitemShut
  {NoStop}%
\bibitem [{\citenamefont {Floerchinger}\ and\ \citenamefont
  {Wetterich}(2009)}]{Floerchinger:2009uf}%
  \BibitemOpen
  \bibfield  {author} {\bibinfo {author} {\bibfnamefont {S.}~\bibnamefont
  {Floerchinger}}\ and\ \bibinfo {author} {\bibfnamefont {C.}~\bibnamefont
  {Wetterich}},\ }\href {\doibase 10.1016/j.physletb.2009.09.014} {\bibfield
  {journal} {\bibinfo  {journal} {Phys.Lett.}\ }\textbf {\bibinfo {volume}
  {B680}},\ \bibinfo {pages} {371} (\bibinfo {year} {2009})},\ \Eprint
  {http://arxiv.org/abs/0905.0915} {arXiv:0905.0915 [hep-th]} \BibitemShut
  {NoStop}%
\bibitem [{\citenamefont {Fischer}\ \emph {et~al.}(2009)\citenamefont
  {Fischer}, \citenamefont {Maas},\ and\ \citenamefont
  {Pawlowski}}]{Fischer:2009ah}%
  \BibitemOpen
  \bibfield  {author} {\bibinfo {author} {\bibfnamefont {C.~S.}\ \bibnamefont
  {Fischer}}, \bibinfo {author} {\bibfnamefont {A.}~\bibnamefont {Maas}}, \
  and\ \bibinfo {author} {\bibfnamefont {J.~M.}\ \bibnamefont {Pawlowski}},\
  }\href {\doibase 10.1016/j.aop.2009.07.009} {\bibfield  {journal} {\bibinfo
  {journal} {Annals Phys.}\ }\textbf {\bibinfo {volume} {324}},\ \bibinfo
  {pages} {2408} (\bibinfo {year} {2009})},\ \Eprint
  {http://arxiv.org/abs/0810.1987} {arXiv:0810.1987 [hep-ph]} \BibitemShut
  {NoStop}%
\bibitem [{\citenamefont {Scavenius}\ \emph {et~al.}(2002)\citenamefont
  {Scavenius}, \citenamefont {Dumitru},\ and\ \citenamefont
  {Lenaghan}}]{Scavenius:2002ru}%
  \BibitemOpen
  \bibfield  {author} {\bibinfo {author} {\bibfnamefont {O.}~\bibnamefont
  {Scavenius}}, \bibinfo {author} {\bibfnamefont {A.}~\bibnamefont {Dumitru}},
  \ and\ \bibinfo {author} {\bibfnamefont {J.}~\bibnamefont {Lenaghan}},\
  }\href {\doibase 10.1103/PhysRevC.66.034903} {\bibfield  {journal} {\bibinfo
  {journal} {Phys.Rev.}\ }\textbf {\bibinfo {volume} {C66}},\ \bibinfo {pages}
  {034903} (\bibinfo {year} {2002})},\ \Eprint
  {http://arxiv.org/abs/hep-ph/0201079} {arXiv:hep-ph/0201079} \BibitemShut
  {NoStop}%
\bibitem [{\citenamefont {Lo}\ \emph {et~al.}(2013)\citenamefont {Lo},
  \citenamefont {Friman}, \citenamefont {Kaczmarek}, \citenamefont {Redlich},\
  and\ \citenamefont {Sasaki}}]{Lo:2013hla}%
  \BibitemOpen
  \bibfield  {author} {\bibinfo {author} {\bibfnamefont {P.~M.}\ \bibnamefont
  {Lo}}, \bibinfo {author} {\bibfnamefont {B.}~\bibnamefont {Friman}}, \bibinfo
  {author} {\bibfnamefont {O.}~\bibnamefont {Kaczmarek}}, \bibinfo {author}
  {\bibfnamefont {K.}~\bibnamefont {Redlich}}, \ and\ \bibinfo {author}
  {\bibfnamefont {C.}~\bibnamefont {Sasaki}},\ }\href@noop {} {\bibfield
  {journal} {\bibinfo  {journal} {Phys. Rev. D 88,}\ }\textbf {\bibinfo
  {volume} {074502}} (\bibinfo {year} {2013})},\ \Eprint
  {http://arxiv.org/abs/1307.5958} {arXiv:1307.5958 [hep-lat]} \BibitemShut
  {NoStop}%
\bibitem [{\citenamefont {Braun}\ \emph
  {et~al.}(2010{\natexlab{b}})\citenamefont {Braun}, \citenamefont {Eichhorn},
  \citenamefont {Gies},\ and\ \citenamefont {Pawlowski}}]{Braun:2010cy}%
  \BibitemOpen
  \bibfield  {author} {\bibinfo {author} {\bibfnamefont {J.}~\bibnamefont
  {Braun}}, \bibinfo {author} {\bibfnamefont {A.}~\bibnamefont {Eichhorn}},
  \bibinfo {author} {\bibfnamefont {H.}~\bibnamefont {Gies}}, \ and\ \bibinfo
  {author} {\bibfnamefont {J.~M.}\ \bibnamefont {Pawlowski}},\ }\href {\doibase
  10.1140/epjc/s10052-010-1485-1} {\bibfield  {journal} {\bibinfo  {journal}
  {Eur.Phys.J.}\ }\textbf {\bibinfo {volume} {C70}},\ \bibinfo {pages} {689}
  (\bibinfo {year} {2010}{\natexlab{b}})},\ \Eprint
  {http://arxiv.org/abs/1007.2619} {arXiv:1007.2619 [hep-ph]} \BibitemShut
  {NoStop}%
\bibitem [{\citenamefont {Ellwanger}\ and\ \citenamefont
  {Wetterich}(1994)}]{Ellwanger:1994wy}%
  \BibitemOpen
  \bibfield  {author} {\bibinfo {author} {\bibfnamefont {U.}~\bibnamefont
  {Ellwanger}}\ and\ \bibinfo {author} {\bibfnamefont {C.}~\bibnamefont
  {Wetterich}},\ }\href {\doibase 10.1016/0550-3213(94)90568-1} {\bibfield
  {journal} {\bibinfo  {journal} {Nucl.Phys.}\ }\textbf {\bibinfo {volume}
  {B423}},\ \bibinfo {pages} {137} (\bibinfo {year} {1994})},\ \Eprint
  {http://arxiv.org/abs/hep-ph/9402221} {arXiv:hep-ph/9402221} \BibitemShut
  {NoStop}%
\bibitem [{\citenamefont {Jungnickel}\ and\ \citenamefont
  {Wetterich}(1996{\natexlab{a}})}]{Jungnickel:1995fp}%
  \BibitemOpen
  \bibfield  {author} {\bibinfo {author} {\bibfnamefont {D.}~\bibnamefont
  {Jungnickel}}\ and\ \bibinfo {author} {\bibfnamefont {C.}~\bibnamefont
  {Wetterich}},\ }\href {\doibase 10.1103/PhysRevD.53.5142} {\bibfield
  {journal} {\bibinfo  {journal} {Phys.Rev.}\ }\textbf {\bibinfo {volume}
  {D53}},\ \bibinfo {pages} {5142} (\bibinfo {year} {1996}{\natexlab{a}})},\
  \Eprint {http://arxiv.org/abs/hep-ph/9505267} {arXiv:hep-ph/9505267}
  \BibitemShut {NoStop}%
\bibitem [{\citenamefont {Schaefer}\ and\ \citenamefont
  {Pirner}(1997)}]{Schaefer:1997nd}%
  \BibitemOpen
  \bibfield  {author} {\bibinfo {author} {\bibfnamefont {B.-J.}\ \bibnamefont
  {Schaefer}}\ and\ \bibinfo {author} {\bibfnamefont {H.-J.}\ \bibnamefont
  {Pirner}},\ }\href {\doibase 10.1016/S0375-9474(97)00601-5} {\bibfield
  {journal} {\bibinfo  {journal} {Nucl.Phys.}\ }\textbf {\bibinfo {volume}
  {A627}},\ \bibinfo {pages} {481} (\bibinfo {year} {1997})},\ \Eprint
  {http://arxiv.org/abs/hep-ph/9706258} {arXiv:hep-ph/9706258} \BibitemShut
  {NoStop}%
\bibitem [{\citenamefont {Berges}\ \emph {et~al.}(1999)\citenamefont {Berges},
  \citenamefont {Jungnickel},\ and\ \citenamefont {Wetterich}}]{Berges:1997eu}%
  \BibitemOpen
  \bibfield  {author} {\bibinfo {author} {\bibfnamefont {J.}~\bibnamefont
  {Berges}}, \bibinfo {author} {\bibfnamefont {D.}~\bibnamefont {Jungnickel}},
  \ and\ \bibinfo {author} {\bibfnamefont {C.}~\bibnamefont {Wetterich}},\
  }\href {\doibase 10.1103/PhysRevD.59.034010} {\bibfield  {journal} {\bibinfo
  {journal} {Phys.Rev.}\ }\textbf {\bibinfo {volume} {D59}},\ \bibinfo {pages}
  {034010} (\bibinfo {year} {1999})},\ \Eprint
  {http://arxiv.org/abs/hep-ph/9705474} {arXiv:hep-ph/9705474} \BibitemShut
  {NoStop}%
\bibitem [{\citenamefont {Schaefer}\ and\ \citenamefont
  {Wambach}(2005)}]{Schaefer:2004en}%
  \BibitemOpen
  \bibfield  {author} {\bibinfo {author} {\bibfnamefont {B.-J.}\ \bibnamefont
  {Schaefer}}\ and\ \bibinfo {author} {\bibfnamefont {J.}~\bibnamefont
  {Wambach}},\ }\href {\doibase 10.1016/j.nuclphysa.2005.04.012} {\bibfield
  {journal} {\bibinfo  {journal} {Nucl.Phys.}\ }\textbf {\bibinfo {volume}
  {A757}},\ \bibinfo {pages} {479} (\bibinfo {year} {2005})},\ \Eprint
  {http://arxiv.org/abs/nucl-th/0403039} {arXiv:nucl-th/0403039} \BibitemShut
  {NoStop}%
\bibitem [{\citenamefont {Lenaghan}\ \emph {et~al.}(2000)\citenamefont
  {Lenaghan}, \citenamefont {Rischke},\ and\ \citenamefont
  {Schaffner-Bielich}}]{Lenaghan:2000ey}%
  \BibitemOpen
  \bibfield  {author} {\bibinfo {author} {\bibfnamefont {J.~T.}\ \bibnamefont
  {Lenaghan}}, \bibinfo {author} {\bibfnamefont {D.~H.}\ \bibnamefont
  {Rischke}}, \ and\ \bibinfo {author} {\bibfnamefont {J.}~\bibnamefont
  {Schaffner-Bielich}},\ }\href {\doibase 10.1103/PhysRevD.62.085008}
  {\bibfield  {journal} {\bibinfo  {journal} {Phys.Rev.}\ }\textbf {\bibinfo
  {volume} {D62}},\ \bibinfo {pages} {085008} (\bibinfo {year} {2000})},\
  \Eprint {http://arxiv.org/abs/nucl-th/0004006} {arXiv:nucl-th/0004006}
  \BibitemShut {NoStop}%
\bibitem [{\citenamefont {Schaefer}\ and\ \citenamefont
  {Wagner}(2009)}]{Schaefer:2008hk}%
  \BibitemOpen
  \bibfield  {author} {\bibinfo {author} {\bibfnamefont {B.-J.}\ \bibnamefont
  {Schaefer}}\ and\ \bibinfo {author} {\bibfnamefont {M.}~\bibnamefont
  {Wagner}},\ }\href {\doibase 10.1103/PhysRevD.79.014018} {\bibfield
  {journal} {\bibinfo  {journal} {Phys.Rev.}\ }\textbf {\bibinfo {volume}
  {D79}},\ \bibinfo {pages} {014018} (\bibinfo {year} {2009})},\ \Eprint
  {http://arxiv.org/abs/0808.1491} {arXiv:0808.1491 [hep-ph]} \BibitemShut
  {NoStop}%
\bibitem [{\citenamefont {Jungnickel}\ and\ \citenamefont
  {Wetterich}(1996{\natexlab{b}})}]{Jungnickel1996b}%
  \BibitemOpen
  \bibfield  {author} {\bibinfo {author} {\bibfnamefont {D.~U.}\ \bibnamefont
  {Jungnickel}}\ and\ \bibinfo {author} {\bibfnamefont {C.}~\bibnamefont
  {Wetterich}},\ }\href {\doibase 10.1103/PhysRevD.53.5142} {\bibfield
  {journal} {\bibinfo  {journal} {Phys. Rev.}\ }\textbf {\bibinfo {volume}
  {D53}},\ \bibinfo {pages} {5142} (\bibinfo {year} {1996}{\natexlab{b}})},\
  \Eprint {http://arxiv.org/abs/hep-ph/9505267} {arXiv:hep-ph/9505267}
  \BibitemShut {NoStop}%
\bibitem [{\citenamefont {'t~Hooft}(1976{\natexlab{a}})}]{Hooft:1976fv}%
  \BibitemOpen
  \bibfield  {author} {\bibinfo {author} {\bibfnamefont {G.}~\bibnamefont
  {'t~Hooft}},\ }\href {\doibase 10.1103/PhysRevD.18.2199.3,
  10.1103/PhysRevD.14.3432} {\bibfield  {journal} {\bibinfo  {journal}
  {Phys.Rev.}\ }\textbf {\bibinfo {volume} {D14}},\ \bibinfo {pages} {3432}
  (\bibinfo {year} {1976}{\natexlab{a}})}\BibitemShut {NoStop}%
\bibitem [{\citenamefont {'t~Hooft}(1976{\natexlab{b}})}]{Hooft:1976up}%
  \BibitemOpen
  \bibfield  {author} {\bibinfo {author} {\bibfnamefont {G.}~\bibnamefont
  {'t~Hooft}},\ }\href {\doibase 10.1103/PhysRevLett.37.8} {\bibfield
  {journal} {\bibinfo  {journal} {Phys.Rev.Lett.}\ }\textbf {\bibinfo {volume}
  {37}},\ \bibinfo {pages} {8} (\bibinfo {year}
  {1976}{\natexlab{b}})}\BibitemShut {NoStop}%
\bibitem [{\citenamefont {Kobayashi}\ \emph {et~al.}(1971)\citenamefont
  {Kobayashi}, \citenamefont {Kondo},\ and\ \citenamefont
  {Maskawa}}]{Kobayashi:1971qz}%
  \BibitemOpen
  \bibfield  {author} {\bibinfo {author} {\bibfnamefont {M.}~\bibnamefont
  {Kobayashi}}, \bibinfo {author} {\bibfnamefont {H.}~\bibnamefont {Kondo}}, \
  and\ \bibinfo {author} {\bibfnamefont {T.}~\bibnamefont {Maskawa}},\ }\href
  {\doibase 10.1143/PTP.45.1955} {\bibfield  {journal} {\bibinfo  {journal}
  {Prog.Theor.Phys.}\ }\textbf {\bibinfo {volume} {45}},\ \bibinfo {pages}
  {1955} (\bibinfo {year} {1971})}\BibitemShut {NoStop}%
\bibitem [{\citenamefont {'t~Hooft}(1986)}]{Hooft:1986nc}%
  \BibitemOpen
  \bibfield  {author} {\bibinfo {author} {\bibfnamefont {G.}~\bibnamefont
  {'t~Hooft}},\ }\href {\doibase 10.1016/0370-1573(86)90117-1} {\bibfield
  {journal} {\bibinfo  {journal} {Phys.Rept.}\ }\textbf {\bibinfo {volume}
  {142}},\ \bibinfo {pages} {357} (\bibinfo {year} {1986})}\BibitemShut
  {NoStop}%
\bibitem [{\citenamefont {Schaefer}\ and\ \citenamefont
  {Mitter}(2013)}]{Schaefer:2013isa}%
  \BibitemOpen
  \bibfield  {author} {\bibinfo {author} {\bibfnamefont {B.-J.}\ \bibnamefont
  {Schaefer}}\ and\ \bibinfo {author} {\bibfnamefont {M.}~\bibnamefont
  {Mitter}},\ }\href@noop {} {\  (\bibinfo {year} {2013})},\ \Eprint
  {http://arxiv.org/abs/1312.3850} {arXiv:1312.3850 [hep-ph]} \BibitemShut
  {NoStop}%
\bibitem [{\citenamefont {Beringer}\ \emph {et~al.}(2012)\citenamefont
  {Beringer} \emph {et~al.}}]{Beringer:1900zz}%
  \BibitemOpen
  \bibfield  {author} {\bibinfo {author} {\bibfnamefont {J.}~\bibnamefont
  {Beringer}} \emph {et~al.} (\bibinfo {collaboration} {Particle Data Group}),\
  }\href {\doibase 10.1103/PhysRevD.86.010001} {\bibfield  {journal} {\bibinfo
  {journal} {Phys.Rev.}\ }\textbf {\bibinfo {volume} {D86}},\ \bibinfo {pages}
  {010001} (\bibinfo {year} {2012})}\BibitemShut {NoStop}%
\bibitem [{\citenamefont {Braun}\ \emph {et~al.}(2004)\citenamefont {Braun},
  \citenamefont {Schwenzer},\ and\ \citenamefont {Pirner}}]{Braun:2003ii}%
  \BibitemOpen
  \bibfield  {author} {\bibinfo {author} {\bibfnamefont {J.}~\bibnamefont
  {Braun}}, \bibinfo {author} {\bibfnamefont {K.}~\bibnamefont {Schwenzer}}, \
  and\ \bibinfo {author} {\bibfnamefont {H.-J.}\ \bibnamefont {Pirner}},\
  }\href {\doibase 10.1103/PhysRevD.70.085016} {\bibfield  {journal} {\bibinfo
  {journal} {Phys.Rev.}\ }\textbf {\bibinfo {volume} {D70}},\ \bibinfo {pages}
  {085016} (\bibinfo {year} {2004})},\ \Eprint
  {http://arxiv.org/abs/hep-ph/0312277} {arXiv:hep-ph/0312277} \BibitemShut
  {NoStop}%
\bibitem [{\citenamefont {Borsanyi}\ \emph
  {et~al.}(2010{\natexlab{a}})\citenamefont {Borsanyi} \emph
  {et~al.}}]{Borsanyi:2010bp}%
  \BibitemOpen
  \bibfield  {author} {\bibinfo {author} {\bibfnamefont {S.}~\bibnamefont
  {Borsanyi}} \emph {et~al.} (\bibinfo {collaboration} {Wuppertal-Budapest
  Collaboration}),\ }\href {\doibase 10.1007/JHEP09(2010)073} {\bibfield
  {journal} {\bibinfo  {journal} {JHEP}\ }\textbf {\bibinfo {volume} {1009}},\
  \bibinfo {pages} {073} (\bibinfo {year} {2010}{\natexlab{a}})},\ \Eprint
  {http://arxiv.org/abs/1005.3508} {arXiv:1005.3508 [hep-lat]} \BibitemShut
  {NoStop}%
\bibitem [{\citenamefont {Strodthoff}\ \emph {et~al.}(2012)\citenamefont
  {Strodthoff}, \citenamefont {Schaefer},\ and\ \citenamefont {von
  Smekal}}]{Strodthoff:2011tz}%
  \BibitemOpen
  \bibfield  {author} {\bibinfo {author} {\bibfnamefont {N.}~\bibnamefont
  {Strodthoff}}, \bibinfo {author} {\bibfnamefont {B.-J.}\ \bibnamefont
  {Schaefer}}, \ and\ \bibinfo {author} {\bibfnamefont {L.}~\bibnamefont {von
  Smekal}},\ }\href {\doibase 10.1103/PhysRevD.85.074007} {\bibfield  {journal}
  {\bibinfo  {journal} {Phys.Rev.}\ }\textbf {\bibinfo {volume} {D85}},\
  \bibinfo {pages} {074007} (\bibinfo {year} {2012})},\ \Eprint
  {http://arxiv.org/abs/1112.5401} {arXiv:1112.5401 [hep-ph]} \BibitemShut
  {NoStop}%
\bibitem [{\citenamefont {Bazavov}(2013)}]{Bazavov:2012bp}%
  \BibitemOpen
  \bibfield  {author} {\bibinfo {author} {\bibfnamefont {A.}~\bibnamefont
  {Bazavov}} (\bibinfo {collaboration} {HotQCD Collaboration}),\ }\href
  {\doibase 10.1016/j.nuclphysa.2013.02.155} {\bibfield  {journal} {\bibinfo
  {journal} {Nucl.Phys.}\ }\textbf {\bibinfo {volume} {A904-905}},\ \bibinfo
  {pages} {877c} (\bibinfo {year} {2013})},\ \Eprint
  {http://arxiv.org/abs/1210.6312} {arXiv:1210.6312 [hep-lat]} \BibitemShut
  {NoStop}%
\bibitem [{\citenamefont {Borsanyi}\ \emph
  {et~al.}(2010{\natexlab{b}})\citenamefont {Borsanyi}, \citenamefont
  {Endrodi}, \citenamefont {Fodor}, \citenamefont {Jakovac}, \citenamefont
  {Katz} \emph {et~al.}}]{Borsanyi:2010cj}%
  \BibitemOpen
  \bibfield  {author} {\bibinfo {author} {\bibfnamefont {S.}~\bibnamefont
  {Borsanyi}}, \bibinfo {author} {\bibfnamefont {G.}~\bibnamefont {Endrodi}},
  \bibinfo {author} {\bibfnamefont {Z.}~\bibnamefont {Fodor}}, \bibinfo
  {author} {\bibfnamefont {A.}~\bibnamefont {Jakovac}}, \bibinfo {author}
  {\bibfnamefont {S.~D.}\ \bibnamefont {Katz}},  \emph {et~al.},\ }\href
  {\doibase 10.1007/JHEP11(2010)077} {\bibfield  {journal} {\bibinfo  {journal}
  {JHEP}\ }\textbf {\bibinfo {volume} {1011}},\ \bibinfo {pages} {077}
  (\bibinfo {year} {2010}{\natexlab{b}})},\ \Eprint
  {http://arxiv.org/abs/1007.2580} {arXiv:1007.2580 [hep-lat]} \BibitemShut
  {NoStop}%
\bibitem [{\citenamefont {Bazavov}\ \emph {et~al.}(2012)\citenamefont
  {Bazavov}, \citenamefont {Bhattacharya}, \citenamefont {Cheng}, \citenamefont
  {DeTar}, \citenamefont {Ding} \emph {et~al.}}]{Bazavov:2011nk}%
  \BibitemOpen
  \bibfield  {author} {\bibinfo {author} {\bibfnamefont {A.}~\bibnamefont
  {Bazavov}}, \bibinfo {author} {\bibfnamefont {T.}~\bibnamefont
  {Bhattacharya}}, \bibinfo {author} {\bibfnamefont {M.}~\bibnamefont {Cheng}},
  \bibinfo {author} {\bibfnamefont {C.}~\bibnamefont {DeTar}}, \bibinfo
  {author} {\bibfnamefont {H.}~\bibnamefont {Ding}},  \emph {et~al.},\ }\href
  {\doibase 10.1103/PhysRevD.85.054503} {\bibfield  {journal} {\bibinfo
  {journal} {Phys.Rev.}\ }\textbf {\bibinfo {volume} {D85}},\ \bibinfo {pages}
  {054503} (\bibinfo {year} {2012})},\ \Eprint {http://arxiv.org/abs/1111.1710}
  {arXiv:1111.1710 [hep-lat]} \BibitemShut {NoStop}%
\bibitem [{\citenamefont {Beringer (Particle Data Group)~et
  al.}(2012)}]{pdg:2012}%
  \BibitemOpen
  \bibfield  {author} {\bibinfo {author} {\bibfnamefont {J.}~\bibnamefont
  {Beringer (Particle Data Group)~et al.}},\ }\href@noop {} {\bibfield
  {journal} {\bibinfo  {journal} {Phys. Rev.}\ }\textbf {\bibinfo {volume}
  {D86}},\ \bibinfo {pages} {010001} (\bibinfo {year} {2012})}\BibitemShut
  {NoStop}%
\bibitem [{\citenamefont {Skokov}\ \emph
  {et~al.}(2010{\natexlab{b}})\citenamefont {Skokov}, \citenamefont {Friman},
  \citenamefont {Nakano}, \citenamefont {Redlich},\ and\ \citenamefont
  {Schaefer}}]{Skokov:2010sf}%
  \BibitemOpen
  \bibfield  {author} {\bibinfo {author} {\bibfnamefont {V.}~\bibnamefont
  {Skokov}}, \bibinfo {author} {\bibfnamefont {B.}~\bibnamefont {Friman}},
  \bibinfo {author} {\bibfnamefont {E.}~\bibnamefont {Nakano}}, \bibinfo
  {author} {\bibfnamefont {K.}~\bibnamefont {Redlich}}, \ and\ \bibinfo
  {author} {\bibfnamefont {B.-J.}\ \bibnamefont {Schaefer}},\ }\href {\doibase
  10.1103/PhysRevD.82.034029} {\bibfield  {journal} {\bibinfo  {journal}
  {Phys.Rev.}\ }\textbf {\bibinfo {volume} {D82}},\ \bibinfo {pages} {034029}
  (\bibinfo {year} {2010}{\natexlab{b}})},\ \Eprint
  {http://arxiv.org/abs/1005.3166} {arXiv:1005.3166 [hep-ph]} \BibitemShut
  {NoStop}%
\bibitem [{\citenamefont {Andersen}\ \emph {et~al.}(2011)\citenamefont
  {Andersen}, \citenamefont {Khan},\ and\ \citenamefont
  {Kyllingstad}}]{Andersen:2011pr}%
  \BibitemOpen
  \bibfield  {author} {\bibinfo {author} {\bibfnamefont {J.~O.}\ \bibnamefont
  {Andersen}}, \bibinfo {author} {\bibfnamefont {R.}~\bibnamefont {Khan}}, \
  and\ \bibinfo {author} {\bibfnamefont {L.~T.}\ \bibnamefont {Kyllingstad}},\
  }\href@noop {} {\  (\bibinfo {year} {2011})},\ \Eprint
  {http://arxiv.org/abs/1102.2779} {arXiv:1102.2779 [hep-ph]} \BibitemShut
  {NoStop}%
\bibitem [{\citenamefont {Gupta}\ and\ \citenamefont
  {Tiwari}(2012)}]{Gupta:2011ez}%
  \BibitemOpen
  \bibfield  {author} {\bibinfo {author} {\bibfnamefont {U.~S.}\ \bibnamefont
  {Gupta}}\ and\ \bibinfo {author} {\bibfnamefont {V.~K.}\ \bibnamefont
  {Tiwari}},\ }\href {\doibase 10.1103/PhysRevD.85.014010} {\bibfield
  {journal} {\bibinfo  {journal} {Phys.Rev.}\ }\textbf {\bibinfo {volume}
  {D85}},\ \bibinfo {pages} {014010} (\bibinfo {year} {2012})},\ \Eprint
  {http://arxiv.org/abs/1107.1312} {arXiv:1107.1312 [hep-ph]} \BibitemShut
  {NoStop}%
\bibitem [{\citenamefont {Marhauser}\ and\ \citenamefont
  {Pawlowski}(2008)}]{Marhauser:2008fz}%
  \BibitemOpen
  \bibfield  {author} {\bibinfo {author} {\bibfnamefont {F.}~\bibnamefont
  {Marhauser}}\ and\ \bibinfo {author} {\bibfnamefont {J.~M.}\ \bibnamefont
  {Pawlowski}},\ }\href@noop {} {\  (\bibinfo {year} {2008})},\ \Eprint
  {http://arxiv.org/abs/0812.1144} {arXiv:0812.1144 [hep-ph]} \BibitemShut
  {NoStop}%
\end{thebibliography}%

\end{document}